\documentclass[a4paper,11pt]{article}
\usepackage{jheppub}
\usepackage{booktabs}
\usepackage{xcolor}
\usepackage{bm}  
\makeatletter
\gdef\@fpheader{}
\makeatother

\title{Thermal Evolution and Hydrodynamic Filtering of Pseudoscalar Dark Matter}

\author[b]{Xiu-Fei Li,}
\author[a]{Hongxin Wang\footnote{Corresponding author}}
\author[a]{Lei Wang}

\affiliation[a]{Department of Physics, Yantai University, Yantai 264005, China}
\affiliation[b]{Inner Mongolia Key Laboratory of Microscale Physics and Atom Manufacturing, School of Physical Science and Technology, Inner Mongolia University, Hohhot 010021, China}

\emailAdd{xiufeili@imu.edu.cn}
\emailAdd{hxwang@ytu.edu.cn}
\emailAdd{leiwang@ytu.edu.cn}

\abstract{
Filtered dark matter provides a mechanism for producing massive dark matter particles during a first order phase transition. The resulting abundance can be modified by plasma hydrodynamics. We investigate this effect in a complex singlet extension of the Standard Model, focusing on the deflagration regime. The entropy normalized dark matter abundance $Y_\chi$ is calculated using both analytic and numerical methods. As the dark matter mass increases, we identify three patterns of abundance evolution before nucleation, ranging from thermal equilibrium to thermal suppression and freeze-out. For $\xi_w=0.01$, the numerical results show that hydrodynamic effects increase the final $Y_\chi$ by factors of approximately $4.1$, $5.7$, and $31.6$ for $m_\chi=1.78$, $2.30$, and $5.03~\mathrm{TeV}$, respectively. For the two lighter masses, the numerical and analytic results agree within $5\%$, and the difference is reduced to about $1\%$ or less when hydrodynamic effects are included. The heaviest case requires a numerical treatment because dark matter freeze-out before nucleation. Our results show that hydrodynamic reheating can enhance $Y_\chi$ even when dark matter has already frozen out before the phase transition. In this regime, $Y_\chi$ is determined by the dark matter abundance established before nucleation and hydrodynamic filtering at the bubble wall.}

\newcommand{\gw}{\gamma_w}
\newcommand{\gp}{\tilde{\gamma}_+}
\newcommand{\vp}{\tilde v_+}
\newcommand{\dm}{\chi}
\newcommand{\rin}{R_{\rm in}}
\newcommand{\dd}{\mathrm{d}}

\begin{document}
\maketitle
\flushbottom

\section{Introduction}

The microscopic origin of dark matter remains one of the central open questions at particle physics and cosmology.  
Thermal freeze-out provides a simple and predictive benchmark\cite{Goldberg:1983nd,Feng:2010gw,Ellis:1983ew,Lee:1977ua,Jungman:1995df,Servant:2002aq,Cheng:2002ej,Bertone:2004pz}, but the absence of confirmed signals in direct, indirect, and collider searches motivates a broader view of dark matter production.  
In particular, if the dark sector undergoes a first order phase transition, the dark matter abundance need not be determined solely by the gradual cooling of the thermal bath\cite{Krylov:2013qe,Huang:2017kzu,Chao:2020adk,Baker:2019ndr,Chway:2019kft,Jiang:2023nkj,Xu:2023lkf,Azatov:2021ifm,Azatov:2022tii,Baldes:2022oev,Xiao:2022oaq}. Unlike a homogeneous and adiabatic thermal evolution, a first order phase transition proceeds through bubble nucleation and wall expansion. The dynamical interface between the false and true vacua can therefore exert additional effects on dark matter, so that the final abundance is not controlled only by the temperature decrease due to cosmic expansion.
Refs.~\cite{Baker:2019ndr,Chway:2019kft} first considered this possibility and proposed the filtered dark matter mechanism.
If the dark matter particle is light in the false vacuum but becomes heavy after crossing the bubble wall, energy conservation implies that only particles with sufficiently large momentum can enter the broken phase.  
The reflected population remains in the symmetric phase and can annihilate away, while the transmitted population survives as the relic dark matter.  
Unlike standard WIMP freeze-out, where the relic abundance is mainly controlled by the annihilation cross section and is therefore subject to the usual unitarity limitation on heavy thermal relics \cite{Baldes:2017gzw,Griest:1989wd,Smirnov:2019ngs}, the filtered dark matter mechanism can suppress the abundance through a Boltzmann like wall transmission factor, opening otherwise inaccessible heavy dark matter parameter space.

The entropy normalized dark matter abundance $Y_\chi$ is exponentially sensitive to the plasma conditions near the bubble wall. In conventional calculations, the temperature and fluid velocity of the incident plasma are often approximated by the nucleation temperature and bubble wall velocity, respectively. These approximations neglect the hydrodynamic response of the plasma to the phase transition. The release of latent heat and the resulting bulk fluid motion modify the temperature and velocity immediately in front of and behind the wall, particularly for strong first order phase transitions \cite{Giese:2020rtr,Giese:2020znk,Leitao:2014pda,Wang:2023jto,Wang:2022lyd,Ai:2021kak,Ai:2023see}. Since the filtering efficiency depends on the incident distribution, these local changes can significantly enhance or suppress the number density. Hydrodynamic effects have been studied extensively in particle transport \cite{Cline:2020jre,Laurent:2020gpg,Cline:2021iff,Lewicki:2021pgr,Wada:2026iux} and gravitational wave production \cite{Si:2025vdt,Tenkanen:2022tly,Tian:2024ysd,Wang:2021jtw}, but the impact on filtered dark matter requires a systematic investigation.

To investigate this hydrodynamic impact, we consider a complex singlet extension of the Standard Model with a softly broken global $U(1)$ symmetry. Such models provide a framework for studying first order phase transitions and Higgs portal dark matter \cite{Profumo:2007wc,Cline:2012hg,Barger:2008jx,Gross:2017dan,Kannike:2019wsn,Kannike:2019mzk,Alanne:2020jwx,Coito:2021fgo}. The stable CP-odd component of the singlet serves as the pseudoscalar dark matter candidate. During a first order phase transition in the singlet direction, its mass changes across the bubble wall, giving rise to the filtering effect. A range of dark matter masses is considered to capture different patterns in the evolution of $Y_\chi$ before the phase transition, including scenarios in which dark matter either remains in equilibrium at nucleation or has already frozen out. For the equilibrium cases, $Y_\chi$ is calculated using both analytic and numerical methods, whereas freeze out before nucleation requires a numerical treatment of the preceding abundance evolution. The analysis focuses on the deflagration regime because the shock front preceding the bubble wall heats and accelerates the plasma before it reaches the wall, making this regime particularly suitable for studying hydrodynamic effects on filtered dark matter. A representative slow wall velocity of \(\xi_w=0.01\) is adopted to quantify the corrections to $Y_\chi$ in each case.

The remainder of this paper is organized as follows. 
Section~\ref{sec:model} summarizes the pseudoscalar dark matter model, the finite temperature effective potential and the benchmark phase transition backgrounds used in the analysis. 
Section~\ref{sec:Analytic} discusses the hydrodynamic modes of the plasma and presents an analytic treatment of the number density for different patterns of dark matter abundance evolution, including hydrodynamic corrections to the filtering process. 
Section~\ref{sec:Numerical} presents the numerical treatment and compares the resulting $Y_\chi$ with the analytic estimates. 
Finally, section~\ref{sec:conclusion} summarizes our conclusions.

\section{Pseudoscalar Dark Matter Model and Phase Transition Background}
\label{sec:model}
We work in a complex singlet pseudoscalar dark matter framework, in which the scalar sector consists of the Standard Model Higgs doublet $H$ and a complex singlet $S$. 
We decompose the scalar fields as
\begin{equation}
	H=
	\begin{pmatrix}
		\phi^+\\
		\left(v_h+\phi+i\eta^0\right)/\sqrt{2}
	\end{pmatrix},
	\qquad
	S=\frac{1}{\sqrt{2}}\left(v_s+s+i\dm\right),
	\label{eq:singlet_decomp}
\end{equation}
The scalar potential is then taken to be 
\begin{align}
	V(H,S) ={}&
	-\mu_h^2 |H|^2-\mu_s^2 |S|^2
	+\lambda_h |H|^4+\lambda_s |S|^4+\lambda_{sh}|H|^2|S|^2
	\nonumber\\
	&-\frac{\mu_2^2}{2}\left(S^2+S^{*2}\right)
	+\frac{\mu_3}{2}\left(S^3+S^{*3}\right)
	+\frac{\mu_4}{2}\left(S^4+S^{*4}\right) .
	\label{eq:tree_potential}
\end{align}
All parameters are taken to be real. The last three terms explicitly break the global $U(1)$ symmetry while preserving the CP transformation $S\to S^*$. Under this transformation, the CP-even field $s$ is even, whereas the CP-odd field $\dm$ is odd. If the vacuum also preserves this symmetry, terms that would induce the decay of a single $\dm$ particle are forbidden. As a result, $\dm$ is stable and can be identified as the dark matter candidate.

In the finite temperature phase transition analysis, we follow the CP-even background fields
\begin{equation}
H\to \frac{1}{\sqrt{2}}
\begin{pmatrix}
	0\\
	\phi_h
\end{pmatrix},
\qquad
S\to \frac{\phi_s}{\sqrt{2}}.
	\label{eq:background_fields}
\end{equation}
The use of high temperature effective potentials follows the standard treatment of finite temperature phase transitions \cite{Dolan:1973qd,Quiros:1999jp,Coleman:1977py,Callan:1977pt,Linde:1981zj}.
\begin{align}
	V_{\rm HT}(\phi_h,\phi_s,T)
	={}&
	-\frac{1}{2}\mu_h^2\phi_h^2
	-\frac{1}{2}\left(\mu_s^2+\mu_2^2\right)\phi_s^2
	+\frac{1}{4}\lambda_h\phi_h^4
	+\frac{1}{4}\left(\lambda_s+\mu_4\right)\phi_s^4
	\nonumber\\
	&+\frac{1}{4}\lambda_{sh}\phi_h^2\phi_s^2
	+\frac{\mu_3}{2\sqrt{2}}\phi_s^3
	+\frac{1}{2}c_h^{\rm th}T^2\phi_h^2
	+\frac{1}{2}c_S^{\rm th}T^2\phi_s^2 .
	\label{eq:highT_potential}
\end{align}
Here the leading thermal coefficients are
\begin{equation}
	c_h^{\rm th}=	
	\frac{9g^2+3g'^2+12y_t^2+4\lambda_{sh}+24\lambda_h}{48},
	\qquad
	c_S^{\rm th}=
	\frac{\lambda_{sh}/2+\lambda_s}{3}.
	\label{eq:thermal_coefficients}
\end{equation}
At zero temperature the CP-even scalar mass matrix is
\begin{equation}
{\cal M}^2=
\begin{pmatrix}
2\lambda_hv_h^2 & \lambda_{sh}v_hv_s\\
\lambda_{sh}v_hv_s &
2\lambda_sv_s^2+\dfrac{3}{2\sqrt{2}}\mu_3v_s+2\mu_4v_s^2
\end{pmatrix}.
\label{eq:scalar_mass_matrix}
\end{equation}
The physical CP-even masses are
\begin{equation}
m_{h,\bm a}^2
=
\frac{1}{2}
\left[
{\cal M}_{\phi \phi}^2+{\cal M}_{ss}^2
\pm
\sqrt{\left({\cal M}_{\phi \phi}^2-{\cal M}_{ss}^2\right)^2
+4\left({\cal M}_{\phi s}^2\right)^2}
\right],
\label{eq:scalar_mass_eigenvalues}
\end{equation}
where \(h\) is identified with the observed Higgs boson and \(\bm a\) is the singlet scalar. 
On a general CP-even background $(\phi_h,\phi_s)$, the field dependent pseudoscalar mass is
\begin{equation}
	m_\dm^2(\phi_h,\phi_s)=	
	-\mu_s^2+\mu_2^2
	+\frac{1}{2}\lambda_{sh}\phi_h^2
	+\left(\lambda_s-3\mu_4\right)\phi_s^2
	-\frac{3}{\sqrt{2}}\mu_3\phi_s .
	\label{eq:mchi_field_dependent}
\end{equation}
At the zero temperature vacuum, after applying the singlet tadpole condition, this reduces to
\begin{equation}
	m_\dm^2=	
	2\mu_2^2-\frac{9}{2\sqrt{2}}\mu_3v_s-4\mu_4v_s^2 .
	\label{eq:mchi_vacuum}
\end{equation}

An important feature of this model is that the pseudoscalar mass need not vanish even when the singlet background is zero. The filtering process therefore generally occurs between two phases with different masses, rather than simply between a massless and a massive state. In particular, the incoming dark matter may have a nonzero false vacuum mass $m_0$.
We first scan the parameter space using CosmoTransitions \cite{Wainwright:2011kj} and select a phase transition background for which the transition proceeds from $(\phi_h,\phi_s)=(0,0)$ to $(0,v_s)$ at the nucleation temperature $T_n=150.54$ GeV. The parameters $m_h$, $m_{\bm a}$, $v_h$, $v_s$, $\mu_3$, $\mu_4$ and $\theta$ are then held fixed, while only the pseudoscalar mass $m_\dm$ is varied. Under this parameterization, varying $m_\dm$ changes the CP-odd mass without modifying the CP-even finite temperature potential or the phase transition background. Three  benchmarks are selected to illustrate different patterns of $Y_\chi$ evolution and decoupling relative to the fixed phase transition. Their input parameters are summarized in Tab.~\ref{tab:benchmarks}.

\begin{table}[t]
	\centering
	\renewcommand{\arraystretch}{1.15}
	\setlength{\tabcolsep}{7pt}
	\begin{tabular}{ c c c c c c }
		\hline
		Benchmark
		& \(m_h\,[\mathrm{GeV}]\)
		& \(m_{\bm a}\,[\mathrm{GeV}]\)
		& \(m_\dm\,[\mathrm{GeV}]\)
		& \(v_s\,[\mathrm{GeV}]\)
		& \(\mu_3\,[\mathrm{GeV}]\) \\
		\hline
		BP1 & \(125.00\) & \(735.20\) & \(1783.81\) & \(511.60\) & \(-1954.65\) \\
		BP2 & \(125.00\) & \(735.20\) & \(2300.30\) & \(511.60\) & \(-1954.65\) \\
		BP3 & \(125.00\) & \(735.20\) & \(5030.30\) & \(511.60\) & \(-1954.65\) \\
		\hline
	\end{tabular}
	\caption{Benchmark points used in this study.}
	\label{tab:benchmarks}
\end{table}

\section{Analytic Calculation of $Y_\chi$ with Hydrodynamic Corrections}
\label{sec:Analytic}

\subsection{Hydrodynamic modes of the plasma}
\label{sec:hydro_modes}

We first summarize the hydrodynamic variables that enter the filtering calculation. The total energy momentum tensor is taken to be
\begin{equation}
T^{\mu\nu}=\omega u^\mu u^\nu-pg^{\mu\nu},
\qquad
\omega=e+p,
\label{eq:fluid_stress_tensor}
\end{equation}
where \(e\), \(p\), and \(\omega\) are the energy density, pressure, and enthalpy density.  Conservation of energy and momentum fluxes across a discontinuous wall gives the two matching conditions
\begin{align}
\omega_+\tilde v_+^2\tilde\gamma_+^2+p_+
&=
\omega_-\tilde v_-^2\tilde\gamma_-^2+p_-,
\nonumber\\
\omega_+\tilde v_+\tilde\gamma_+^2
&=
\omega_-\tilde v_-\tilde\gamma_-^2,
\label{eq:hydro_matching}
\end{align}
with \(\tilde\gamma_\pm=(1-\tilde v_\pm^2)^{-1/2}\).  
For orientation, it is useful to adopt the bag model parametrization used in the hydrodynamic analysis,
\begin{equation}
p_+=\frac{1}{3}a_+T_+^4-\epsilon,
\qquad
e_+=a_+T_+^4+\epsilon,
\qquad
p_-=\frac{1}{3}a_-T_-^4,
\qquad
e_-=a_-T_-^4 ,
\label{eq:bag_eos}
\end{equation}
and to define
\begin{equation}
\alpha_+ \equiv \frac{\epsilon}{a_+T_+^4},
\qquad
r_\omega \equiv \frac{\omega_+}{\omega_-}.
\label{eq:alpha_romega}
\end{equation}
The matching conditions then imply
\begin{equation}
\tilde v_+\tilde v_-
=
\frac{p_+-p_-}{e_+-e_-},
\qquad
\frac{\tilde v_+}{\tilde v_-}
=
\frac{e_-+p_+}{e_++p_-}.
\label{eq:velocity_products}
\end{equation}
Equivalently, for the bag model equation of state one may write
\begin{equation}
\tilde v_+
=
\frac{1}{1+\alpha_+}
\left[
\left(
\frac{\tilde v_-}{2}
+\frac{1}{6\tilde v_-}
\right)
\pm
\sqrt{
\left(
\frac{\tilde v_-}{2}
+\frac{1}{6\tilde v_-}
\right)^2
+\alpha_+^2+\frac{2}{3}\alpha_+-\frac{1}{3}}
\right].
\label{eq:vplus_branch}
\end{equation}
The two signs correspond to the detonation and deflagration branches. The velocity in the bubble center frame is related to the wall frame velocity through the Lorentz transformation
\begin{equation}
v_\pm=\mu(\xi_w,\tilde v_\pm),
\qquad
\mu(\xi,v)=\frac{\xi-v}{1-\xi v}.
\label{eq:mu_lorentz}
\end{equation}

For a given bubble wall velocity $\xi_w$, the hydrodynamic solution can in general correspond to a deflagration, detonation, or hybrid configuration \cite{Steinhardt:1981ct,Espinosa:2010hh,Kurki-Suonio:1995rrv,Kurki-Suonio:1995yaf}. In this work, we focus on the deflagration regime, because only in this case does a shock front propagate ahead of the wall and modify the temperature and fluid velocity of the plasma incident on the wall. We therefore summarize only the properties of the deflagration solution needed in our analysis.

The deflagration solution lies on the lower branch of eq.~\eqref{eq:vplus_branch}. In what follows, $c_s$ denotes the speed of sound, which is close to $1/\sqrt{3}$ for a relativistic plasma. The incoming flow in the wall frame is subsonic,
\begin{equation}
\tilde v_+<c_s,
\qquad
\tilde v_+<\tilde v_- .
\label{eq:deflagration_conditions}
\end{equation}
For a spherical bubble, the fluid behind the wall is approximately at rest in the bubble-center frame, so that $\tilde v_-=\xi_w$. A shock front precedes the bubble wall and heats and accelerates the plasma before it reaches the wall. Consequently, the filtering calculation is controlled by the local quantities $T_+$ and $\tilde v_+$, rather than simply by the nucleation temperature $T_n$ and the wall velocity $\xi_w$. Shock heating increases $T_+$, while $\tilde v_+$ is typically reduced relative to $\xi_w$, these two effects act differently on the boosted Boltzmann distribution and can therefore compete in determining the transmission probability through the wall.

\subsection{Freeze-out histories and the analytic result}
\label{sec:boltzmann}
\label{sec:numerical}

We first describe the thermal evolution of dark matter in the false vacuum and define
\begin{equation}
	Y_\dm \equiv \frac{n_\dm}{s},
	\qquad
	x \equiv \frac{m_\dm}{T},
	\label{eq:yield_definition}
\end{equation}
where
\begin{equation}
	s=\frac{2\pi^2}{45}g_{*s}T^3
\end{equation}
is the entropy density.  In the absence of bubble wall filtering, the thermal evolution is governed by the standard Boltzmann equation
\begin{equation}
	\frac{dY_\dm}{dx}
	=
	-\frac{s\langle\sigma v\rangle}{Hx}
	\left[
	Y_\dm^2-\left(Y_\dm^{\rm eq}\right)^2
	\right],
	\label{eq:boltzmann_yield}
\end{equation}
with
\begin{equation}
	Y_\dm^{\rm eq}
	=
	\frac{45}{4\pi^4}
	\frac{g_\dm}{g_{*s}}
	x^2 K_2(x)
	\label{eq:equilibrium_yield}
\end{equation}
in the Maxwell Boltzmann approximation.  This false vacuum evolution determines the abundance incident on the bubble wall at the phase transition.
\begin{figure}[t]
	\centering
	\includegraphics[width=0.88\textwidth]{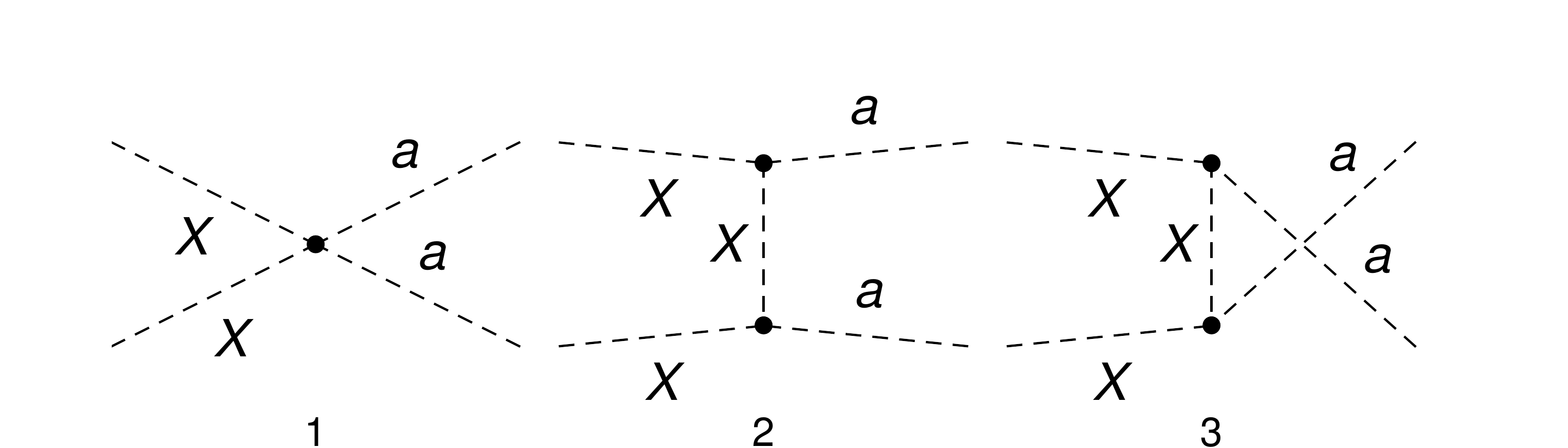}
 	\caption{The tree level diagrams for the annihilation process \(\dm\dm\leftrightarrow aa\).}
	\label{fig:cross_section}
\end{figure}
Before the phase transition, the dark matter particles remain in thermal contact with the plasma through the annihilation process \(\dm\dm\leftrightarrow aa\).  We compute the corresponding thermally averaged annihilation cross section from the tree level diagrams shown in Fig.~\ref{fig:cross_section}. The thermal average is evaluated using the treatment of annihilation processes in the early Universe \cite{Gondolo:1990dk}. Including the symmetry factor for identical final state particles in \(\sigma_{\dm\dm\to aa}\), the thermal average is given by

\begin{align}
	\left\langle \sigma v \right\rangle_{AB\to CD}
	=&
	\frac{1}{8m_A^2m_B^2T K_2(m_A/T)K_2(m_B/T)}
	\nonumber\\
	&\times
	\int_{(m_A+m_B)^2}^{\infty} ds\,
	\frac{\sigma_{AB\to CD}(s)}{\sqrt{s}}\,
	\lambda(s,m_A^2,m_B^2)\,
	K_1\!\left(\frac{\sqrt{s}}{T}\right),
	\label{eq:thermal_average_cross_section}
\end{align}
where
\begin{equation}
	\lambda(s,m_A^2,m_B^2)
	=
	\left[s-(m_A-m_B)^2\right]
	\left[s-(m_A+m_B)^2\right].
	\label{eq:kallen_function}
\end{equation}
For the process of interest, we set \(A=B=\dm\) and \(C=D=\bm a\), using the field dependent masses when the thermal history crosses different background phases. The model is implemented by FeynRules \cite{Alloul:2013bka}, then the decay rates and the scattering cross sections are calculated by FeynCalc\cite{Shtabovenko:2016sxi,Shtabovenko:2020gxv}.

We now turn to the filtering effect induced by the bubble wall.  A dark matter particle incident from the false vacuum side can enter the broken phase only if its wall frame momentum normal to the wall is large enough to compensate for the mass increase,
\begin{equation}
	p_z^w>\sqrt{\Delta m_\dm^2}.
	\label{eq:filter_condition}
\end{equation}
For an equilibrium incoming distribution with false vacuum mass \(m_0\), the distribution in front of the wall is
\begin{equation}
	f_{\dm,+}^{\rm eq}(p_z,p_\perp)
	=
	\left[
	\exp\left(
	\frac{\gp(E_0-\vp p_z)}{T_+}
	\right)\mp 1
	\right]^{-1},
	\qquad
	E_0=\sqrt{p_z^2+p_\perp^2+m_0^2},
	\label{eq:eq_distribution}
\end{equation}
where \(T_+\) and \(\vp\) are the temperature and incoming fluid velocity immediately in front of the wall.  The minus sign applies to bosonic dark matter, while the plus sign applies to
fermionic dark matter.

The transmitted particle flux in the wall frame is
\begin{equation}
	J_\dm^w
	=
	g_\dm
	\int\frac{d^3p}{(2\pi)^3}
	\frac{p_z}{E_0}
	f_{\dm,+}(p_z,p_\perp)
	\Theta\!\left(p_z-\sqrt{\Delta m_\dm^2}\right).
	\label{eq:wall_flux}
\end{equation}
The corresponding number density in the bubble center frame is
\begin{equation}
	n_\dm^{\rm in}=\frac{J_\dm^w}{\gw \xi_w},
	\qquad
	\gw=(1-\xi_w^2)^{-1/2}.
	\label{eq:nin_flux}
\end{equation}
We define the penetration rate as
\begin{equation}
	\rin \equiv \frac{n_\dm^{\rm in}}{n_\dm^{\rm out}},
	\label{eq:penetration_rate}
\end{equation}
where \(n_\dm^{\rm out}\) is evaluated on the false vacuum side.

We define the analytic yield immediately after wall transmission as
\begin{equation}
	Y_{\mathrm{ana}}
	\equiv \frac{n_\dm^{\rm in}}{s_-}
	=\frac{J_\dm^w}{\gamma_w\xi_w\,s_-},
	\qquad
	s_-=\frac{2\pi^2}{45}g_{*s}(T_-)T_-^3.
	\label{eq:Yana_wall}
\end{equation}
This definition is valid for a general false-vacuum mass $m_0$, provided that the incident distribution is described by the equilibrium form in Eq.~\eqref{eq:eq_distribution}. In the Maxwell--Boltzmann approximation, the transverse momentum integral in Eq.~\eqref{eq:wall_flux} can be carried out analytically. One obtains the general expression
\begin{align}
    n_\dm^{\rm in}
    =\frac{g_\dm T_+}{4\pi^2\gw\xi_w\gp}
    \int_{p_{\min}}^\infty dp_z\,p_z
    \exp\left[-\frac{\gp}{T_+}
    \left(\sqrt{p_z^2+m_0^2}-\vp p_z\right)\right].
    \label{eq:nin_general_m0}
\end{align}
For BP2, where $m_0\neq0$, this remaining one-dimensional integral is evaluated numerically. Thus, the label ``analytic'' refers to the equilibrium filtering treatment, while the last momentum integral need not have an elementary closed form for a general $m_0$.

\begin{figure}[!thbp]
	\centering
	\includegraphics[width=0.49\textwidth]{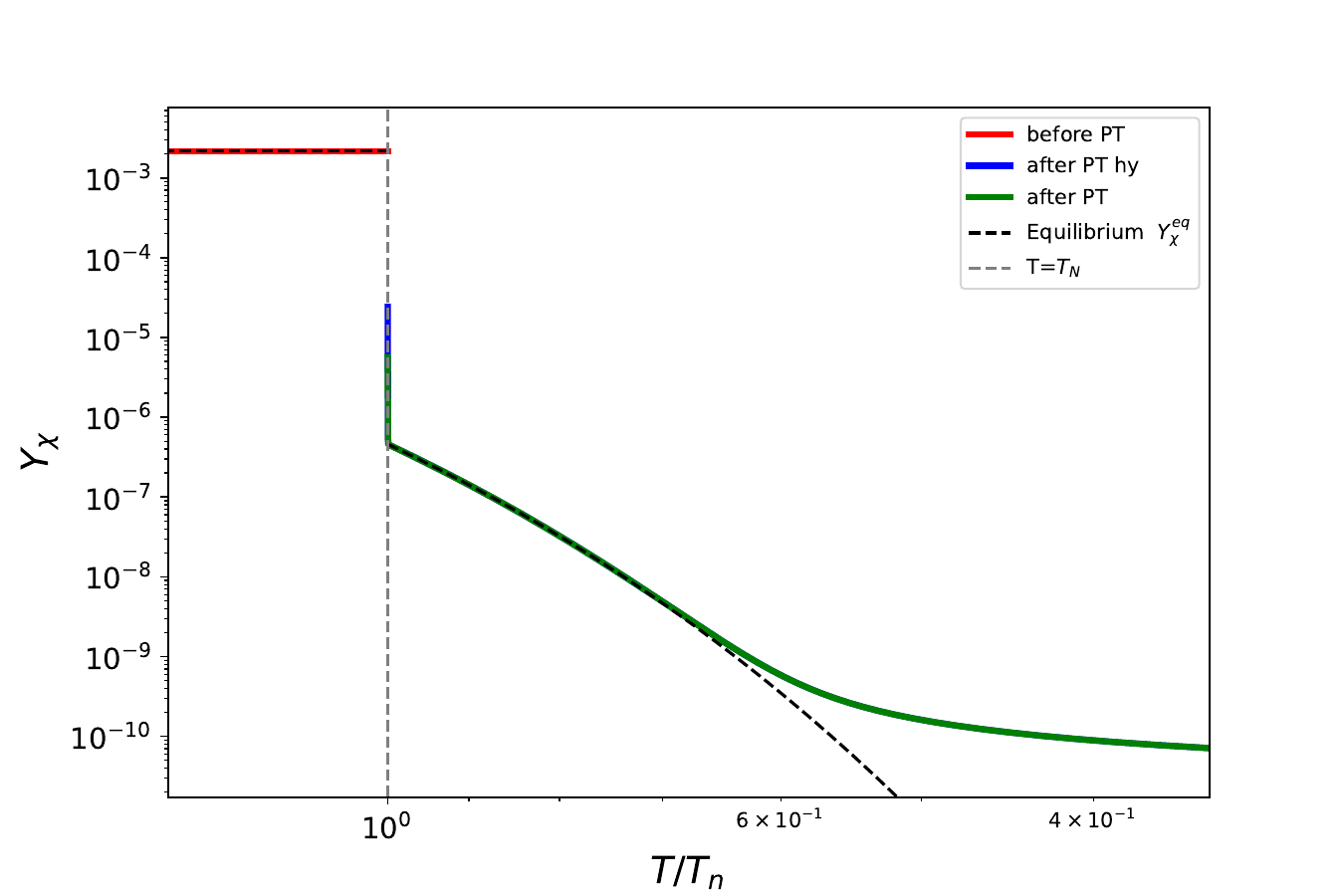} 
	\includegraphics[width=0.49\textwidth]{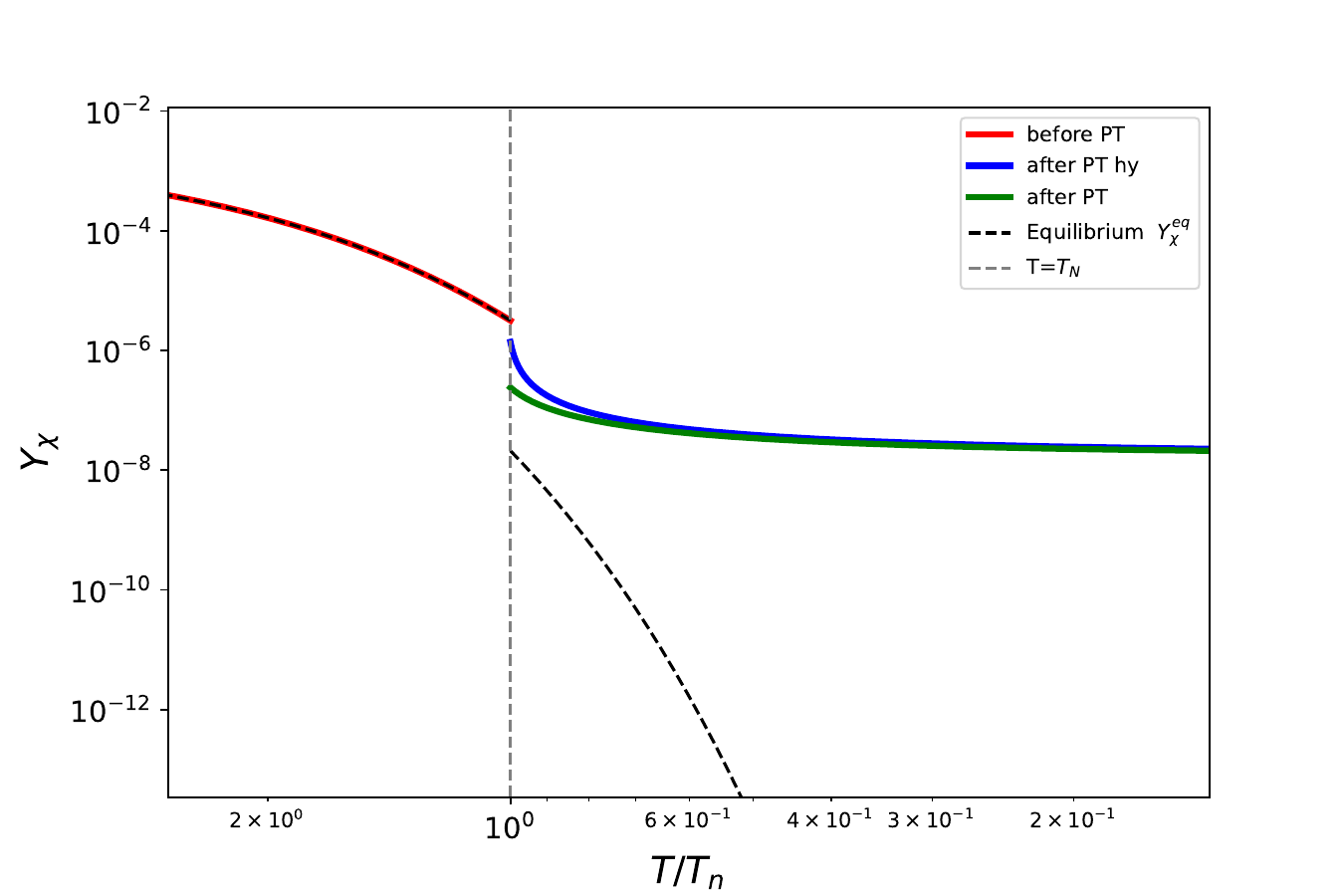} 
	\caption{Dark matter number density as a function of temperature for BP1 (left), BP2 (right). These results are obtained by analytical  methods.}
	\label{fig:analytic_yields}
\end{figure}

The parametric dependence becomes especially transparent in the limiting case $m_0=0$. In this case $p_{\min}=m_\dm^{\rm in}$ and $\sqrt{p_z^2+m_0^2}=p_z$ over the integration region, so Eq.~\eqref{eq:nin_general_m0} can be integrated in closed form, giving
\begin{align}
    n_\dm^{\rm in}
    \simeq
    \frac{g_\dm T_+^3}{4\pi^2\gw \xi_w\,\gp^3(1-\vp)^2}
    \left[
    \gp(1-\vp)\frac{m_\dm^{\rm in}}{T_+}+1
    \right]
    \exp\left[-\gp(1-\vp)\frac{m_\dm^{\rm in}}{T_+}\right].
    \label{eq:nin_hydro_analytic}
\end{align}
Eq.~\eqref{eq:Yana_wall} therefore has the closed form
\begin{align}
	Y_{\mathrm{ana}}
	\simeq
	\frac{45g_\dm}{8\pi^4g_{*s}(T_-)}
	\left(\frac{T_+}{T_-}\right)^3
	\frac{\gp(1-\vp)m_\dm^{\rm in}/T_+ +1}
	{\gw\xi_w\,\gp^3(1-\vp)^2}
	\exp\left[-\gp(1-\vp)\frac{m_\dm^{\rm in}}{T_+}\right].
	\label{eq:Yana_wall_closed}
\end{align}
Taking \(Y_{\mathrm{ana}}\) as the initial condition after wall transmission, the subsequent evolution of the dark matter abundance is obtained by solving Eq.~\eqref{eq:boltzmann_yield}, which gives the evolution curves shown in Fig.~\ref{fig:analytic_yields}.

For a single real scalar degree of freedom, the corresponding relic abundance is
\begin{equation}
	\Omega_\dm h^2
	\simeq
	6.29\times 10^8
	\left(\frac{m_\dm^{\rm in}}{\rm GeV}\right)
	\frac{n_\dm^{\rm in}}{g_{*s}(T_-)T_-^3}.
	\label{eq:omega_hydro}
\end{equation}

Equations~\eqref{eq:nin_hydro_analytic} and~\eqref{eq:omega_hydro} show that hydrodynamic effects can affect the final filtered abundance through two channels.  First,  modify the incoming distribution in front of the
wall. Larger \(T_+\) makes the distribution hotter, allowing more particles to overcome the mass increase across the wall. In addition, the fluid velocity in the wall frame enters the boosted Boltzmann factor,
\begin{equation}
	f_{\dm,+}\propto
	\exp\left[
	-\frac{\tilde\gamma_+\left(E-\tilde v_+p_z\right)}{T_+}
	\right].
	\label{eq:hydro_tail}
\end{equation}
For particles moving toward the wall, \(p_z>0\), the term \(\tilde v_+p_z\) reduces the effective energy \(E-\tilde v_+p_z\), thereby weakening the exponential suppression of the incoming flux.
Second, the transmitted number density alone does not determine the final relic abundance, because $Y_\dm$ is normalized by the entropy density in the broken phase. Since \(s_-\propto g_{*s}(T_-)T_-^3\), an increase in \(T_-\) can dilute the final dark matter $Y_\dm$ even if more particles are transmitted through the wall.  Moreover, the same hydrodynamic solution can also change the broken phase scalar background, thereby modifying \(m_\dm^{\rm in}\) and the dilution factor in Eq.~\eqref{eq:omega_hydro}.  Therefore, the net hydrodynamic correction is not universal, but depends on the detailed thermal and hydrodynamic properties of each benchmark point.

Fig.~\ref{fig:analytic_yields} shows the evolution of $Y_\chi$ as a function of $T/T_n$ for BP1 and BP2. The vertical dashed line marks the phase transition at $T=T_n$. The red curves show the pre-transition evolution, the blue curves show the post-transition $Y_\dm$ including hydrodynamic corrections, and the green curves show the corresponding result without hydrodynamic corrections. The black dashed curves denote the equilibrium $Y_\chi^{\rm eq}$.
Before the phase transition, the two benchmark points exhibit different thermal histories. In the left panel, with
$m_\chi=1783.81~{\rm GeV}$ and $m_0=0$, the dark matter particle is nearly massless in the false vacuum. As a result, the pre-transition $Y_\dm$ remains almost constant, with no significant Boltzmann suppression or freeze-out behavior before nucleation. This means that the incident population is still abundant when the wall forms.
In the right panel, with $m_\chi=2300.30~{\rm GeV}$ and $m_0=1452.38~{\rm GeV}$, the false vacuum mass is already nonzero. $Y_\dm$ decreases gradually before the phase transition, indicating that the incoming population has already been thermally suppressed to some extent before reaching the wall.

At the phase transition, $Y_\dm$ is abruptly affected by the filtering effect of the bubble wall. This change is not an ordinary continuous thermal freeze-out process. Instead, it is caused by the rapid increase of the dark matter mass across
the wall: only particles in the high momentum tail of the incoming distribution can overcome the mass barrier and enter the broken phase. In the left panel, where $m_0=0$, the mass jump across the wall is the largest, and the filtering effect is particularly transparent. $Y_\dm$ is strongly reduced immediately after the phase transition. 
In the right panel, the incoming particles already have a nonzero false vacuum mass, so the post-transition abundance is controlled by both the pre-transition thermal suppression and the wall filtering probability.
After the phase transition, the evolution in the broken phase determines the final relic abundance. In the left panel, $Y_\dm$ continues to decrease after the wall crossing and eventually departs from the equilibrium curve. This indicates that the final abundance is not determined solely by the instantaneous wall filtering, but is also affected by residual annihilation in the broken phase. In the right panel, the post-transition curves quickly become nearly flat, showing that the filtered abundance rapidly freezes to its final value. 

\begin{table}[t]
	\centering
	\begin{tabular}{ c c c c c }
		\hline
		Benchmark & Treatment & \(m_\dm\,[\mathrm{GeV}]\) & \(m_0\,[\mathrm{GeV}]\) & \(Y_{\rm ana}\) \\
		\hline
		BP1 & no hydro. & \(1783.81\) & \(0.00\) & \(5.65\times 10^{-6}\) \\
		BP1 & hydro. & \(1783.81\) & \(0.00\) & \(2.42\times 10^{-5}\) \\
		\hline
		BP2 & no hydro. & \(2300.30\) & \(1452.38\) & \(2.40\times 10^{-7}\) \\
		BP2 & hydro. & \(2300.30\) & \(1452.38\) & \(1.38\times 10^{-6}\) \\
		\hline
	\end{tabular}
	\caption{Analytic wall-filtered yields \(Y_{\rm ana}\) for the benchmark points without and with hydrodynamic effects. Here \(Y_{\rm ana}\equiv n_\chi^{\rm in}/s_-\) denotes the dark matter yield immediately after transmission through the bubble wall.}
	\label{tab:analytic_hy_comparison}
\end{table}

\subsection{Hydrodynamic correction to the analytic estimate}

The impact of hydrodynamic corrections is reflected in the difference between the blue and green curves in Fig.~\ref{fig:analytic_yields}.  These corrections modify the temperature and velocity profiles near the wall and
therefore change the incoming Boltzmann distribution, while leaving the kinematic filtering condition in Eq.~\eqref{eq:filter_condition} unchanged. As summarized in Tab.~\ref{tab:analytic_hy_comparison}, hydrodynamic effects enhance the analytic $Y_\dm$ by factors of approximately 4.3 and 5.8 for BP1 and BP2, respectively.
This mass dependence arises from the exponential behavior in Eq.~\eqref{eq:hydro_tail}.  When \(m_\dm/T_+\) is large, only a small fraction of highly energetic particles can pass through the wall, so small shifts in \(T_+\) or \(\tilde v_+\) are exponentially amplified. This explains why the hydrodynamic correction remains relatively small for the lighter benchmark, where annihilation after the phase transition partially reduces the difference generated at the wall, whereas the correction is much larger for the heavier benchmark.

\begin{figure}[t]
	\centering
	\includegraphics[width=0.62\textwidth]{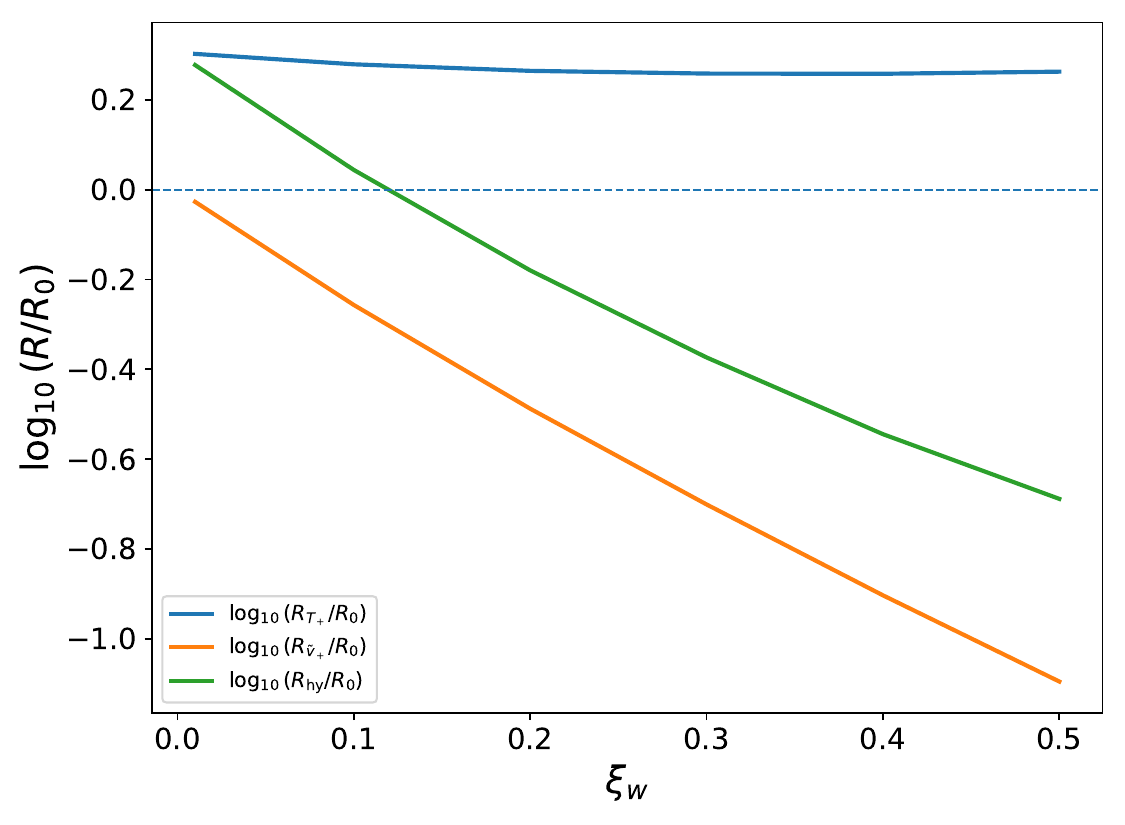}
	\caption{The penetration rate components $R_{T_+}/R_0$, $R_{\tilde v_+}/R_0$ and $R_{\rm hy}/R_0$ as a function of $\xi_w$, respectively.  }
	\label{fig:decomposition_1783}
\end{figure}

To further identify the origin of the hydrodynamic correction, we decompose the penetration rate into the separate effects of the temperature \(T_+\) and the fluid velocity \(\tilde v_+\). In the Maxwell Boltzmann approximation, the penetration rate including hydrodynamic effects is given by
\begin{equation}
R_{\rm hy}
=
\frac{1}{\gamma_w \xi_w}
\frac{
\displaystyle
\tilde{\gamma}_+(1-\tilde v_+)m_\chi^{\rm in}/T_+
+1
}{
4\tilde{\gamma}_+^3(1-\tilde v_+)^2
}
\exp\left[
-\frac{
\tilde{\gamma}_+(1-\tilde v_+)m_\chi^{\rm in}
}{T_+}
\right],
\label{eq:hydrodynamic_penetration_rate}
\end{equation}
For illustration, we focus on the \(m_\dm=1783.81~{\rm GeV}\) benchmark and vary the wall velocity \(\xi_w\). The result of this decomposition is shown in Fig.~\ref{fig:decomposition_1783}. Here \(R_0\) denotes the result without
hydrodynamic corrections, while \(R_{T_+}\) and \(R_{\tilde v_+}\) are obtained by turning on the temperature and velocity corrections separately. The temperature contribution \(R_{T_+}/R_0\) is positive and relatively stable over the range of wall velocities considered, giving an enhancement of roughly \(10^{0.25}\)--\(10^{0.3}\), namely a factor of about two. By contrast, the velocity contribution \(R_{\tilde v_+}/R_0\) decreases as \(\xi_w\) increases.  It is close to unity for \(\xi_w\simeq 0.01\), but becomes an \(\mathcal{O}(10)\) suppression by \(\xi_w\simeq 0.5\). The full hydrodynamic correction is therefore controlled by the competition between these two effects. For the slow-wall benchmark \(\xi_w=0.01\) used in our main analysis, the velocity correction is close to unity and the net enhancement is dominated by shock heating, which increases the high-momentum tail of the incident dark matter distribution. At larger wall velocities, however, the velocity contribution becomes increasingly important and can partially compensate for, or even overtake, the temperature enhancement.

\section{Boltzmann Treatment and Numerical Results}
\label{sec:Numerical}
Although we have generalized the kinematic analysis to the case with a nonzero false vacuum mass \(m_0\), this should be distinguished from the assumption of thermal equilibrium.
The inclusion of \(m_0\neq 0\) modifies the dispersion relation in the false vacuum,
\begin{equation}
	E_\dm^2=p_z^2+p_\perp^2+m_0^2,
\end{equation}
as well as the kinematic transmission condition across the wall.  However, the analytic expression is still obtained by assuming that the incoming distribution in front of the wall is given by the local equilibrium distribution \(f_{\dm,+}^{\rm eq}\).
This assumption is not necessarily valid in the situation we interest.  In particular, the pseudoscalar dark matter may chemically decouple in the false vacuum before the first order phase transition.
In this case, $Y_\dm$ at the nucleation temperature \(T_n\) is no longer fixed by the equilibrium density at the wall. Therefore, the equilibrium analytic estimate has to be generalized to a non-equilibrium treatment.
Similar semiclassical Boltzmann and transport approaches have been extensively used in the study of particle transport and bubble wall dynamics during first order phase transitions \cite{Joyce:1994zn,Moore:1995si,Cline:2000nw}.

The dark matter distribution as
\begin{equation}
	f_\dm(z,p_z,p_\perp)
	=
	\mathcal A(z,p_z)\,
	f_{\dm,+}^{\rm eq}(z,p_z,p_\perp).
	\label{eq:A_definition}
\end{equation}
Here \(f_{\dm,+}^{\rm eq}\) denotes the reference equilibrium distribution in the false vacuum phase, while \(\mathcal A(z,p_z)\) encodes the departure from thermal equilibrium.  The equilibrium result is recovered in the limit \(\mathcal A=1\).  For \(\mathcal A\neq 1\), the incident abundance is instead determined by the solution of the Boltzmann equation,
\begin{equation}
	\bm L[f_\dm]=\bm C[f_\dm],
	\label{eq:boltzmann_general}
\end{equation}
where \(\bm L[f_\dm]\) is the Liouville operator and \(\bm C[f_\dm]\) is the collision operator. 

\subsection{Liouville operator}

For a stationary planar wall, the distribution function has no explicit time dependence and depends on the spatial coordinate only through \(z\). In the wall frame, the single particle energy is conserved,
\begin{equation}
	E^2=p_\perp^2+p_z^2+m_\dm^2(z).
\end{equation}
The corresponding Hamiltonian equations give
\begin{equation}
	\dot z=\frac{p_z}{E},
	\qquad
	\dot p_z
	=
	-\frac{1}{2E}\frac{\dd m_\dm^2(z)}{\dd z}.
\end{equation}
Therefore, the Liouville operator becomes
\begin{equation}
	\bm{L}[f_\dm]
	=
	\frac{p_z}{E}\frac{\partial f_\dm}{\partial z}
	-
	\frac{1}{2E}
	\frac{\dd m_\dm^2(z)}{\dd z}
	\frac{\partial f_\dm}{\partial p_z}.
	\label{eq:liouville_operator_basic}
\end{equation}
Now substitute
\begin{equation}
	f_\dm
	=
	\mathcal A(z,p_z)
	\exp\left[-\frac{E_+^P}{T_+}\right],
	\qquad
	E_+^P=\gp(E-\vp p_z).
	\label{eq:f_substitution}
\end{equation}
The required derivatives are
\begin{align}
	\frac{\partial f_\dm}{\partial z}
	&=
	\exp\left[-\frac{E_+^P}{T_+}\right]
	\left[
	\frac{\partial \mathcal A}{\partial z}
	-
	\frac{\mathcal A}{T_+}
	\frac{\partial E_+^P}{\partial z}
	\right],
	\label{eq:dfdz}
	\\
	\frac{\partial f_\dm}{\partial p_z}
	&=
	\exp\left[-\frac{E_+^P}{T_+}\right]
	\left[
	\frac{\partial \mathcal A}{\partial p_z}
	-
	\frac{\mathcal A}{T_+}
	\frac{\partial E_+^P}{\partial p_z}
	\right].
	\label{eq:dfdpz}
\end{align}
At fixed \(p_z\) and \(p_\perp\),
\begin{equation}
	\frac{\partial E}{\partial z}
	=
	\frac{m_\dm m_\dm'}{E},
	\qquad
	\frac{\partial E}{\partial p_z}
	=
	\frac{p_z}{E}.
	\label{eq:E_derivatives}
\end{equation}
where \(m_\chi'\equiv dm_\chi/dz\).  Hence
\begin{align}
	\frac{\partial E_+^P}{\partial z}
	&=
	\gp
	\frac{m_\dm m_\dm'}{E},
	\label{eq:Eplus_dz}
	\\
	\frac{\partial E_+^P}{\partial p_z}
	&=
	\gp
	\left(
	\frac{p_z}{E}
	-
	\vp
	\right).
	\label{eq:Eplus_dpz}
\end{align}
Using Eqs.~\eqref{eq:dfdz}--\eqref{eq:Eplus_dpz}, we find
\begin{align}
	\bm{L}[f_\dm]
	=
	\frac{\exp\left[-E_+^P/T_+\right]}{E}
	\left[
	p_z\frac{\partial \mathcal A}{\partial z}
	-
	m_\dm m_\dm'\frac{\partial \mathcal A}{\partial p_z}
	-
	\frac{\gp\vp}{T_+}m_\dm m_\dm' \mathcal A
	\right].
	\label{eq:liouville_unintegrated}
\end{align}

To obtain the one dimensional equation in \(z\) and \(p_z\), we integrate over the transverse momenta and multiply by the number of internal degrees of freedom \(g_\dm\),
\begin{equation}
	g_\dm
	\int
	\frac{\dd p_x\,\dd p_y}{(2\pi)^2}
	\bm{L}[f_\dm].
	\label{eq:IL_def}
\end{equation}
Then
\begin{align}
	\int
	\frac{\dd p_x\,\dd p_y}{(2\pi)^2}
	\frac{\exp[-E_+^P/T_+]}{E}
	=
	\frac{T_+}{2\pi\gp}
	\exp\left[
	\frac{\gp(\vp p_z-\sqrt{p_z^2+m_\dm^2(z)})}{T_+}
	\right].
	\label{eq:transverse_integral}
\end{align}
Therefore
\begin{align}
	g_\dm
	\int
	\frac{\dd p_x\,\dd p_y}{(2\pi)^2}
	\bm{L}[f_\dm]
	=&
	\left[
	\left(
	\frac{p_z}{m_\dm}
	\frac{\partial}{\partial z}
	-
	m_\dm'
	\frac{\partial}{\partial p_z}
	-
	m_\dm'
	\frac{\gp\vp}{T_+}
	\right)
	\mathcal A(z,p_z)
	\right]
	\frac{g_\dm m_\dm T_+}{2\pi\gp}\notag \\ 
	&\quad\quad\times \exp\left[
	\frac{\gp(\vp p_z-\sqrt{p_z^2+m_\dm^2(z)})}{T_+}
	\right].
	\label{eq_liouville}
\end{align}

\subsection{Collision Term for \(\chi\chi \leftrightarrow aa\)}

We now derive the collision term used in the Boltzmann equation. The collision term can be evaluated in the plasma frame, and then boosted back to the bubble wall frame. In the present pseudoscalar dark matter model, the relevant annihilation process is
\begin{equation}
	\chi(p)+\chi(q)
	\leftrightarrow
	\bm a(k)+\bm a(l) ,
\end{equation}
Here \(p,q,k,l\) denote the four momenta of the corresponding particles. Explicitly,
\begin{equation}
	p^\mu=(E_p,\mathbf p),\qquad
	q^\mu=(E_q,\mathbf q),\qquad
	k^\mu=(E_k,\mathbf k),\qquad
	l^\mu=(E_l,\mathbf l).
\end{equation}
For a generic on-shell particle with four momentum \(P_i^\mu=(E_i,\mathbf P_i)\), we use the Lorentz invariant phase space
measure
\begin{equation}
	d\Pi_i
	\equiv
	\frac{d^3 \mathbf P_i}{(2\pi)^3 2E_i},
	\qquad
	E_i
	=
	\sqrt{\mathbf P_i^2+m_i^2}.
\end{equation}

For a single real scalar dark matter degree of freedom one should take \(g_\chi=1\). We keep \(g_\chi\) explicitly below in order to make the phase space normalization transparent. After integrating over the transverse momenta of the tagged \(\chi\) particle, the collision term is
\begin{align}
	g_\chi
	\int
	\frac{dp_x dp_y}{(2\pi)^2}
	\mathbf C[f_\chi]
	&=
	-g_\chi^2
	\int
	\frac{dp_x dp_y}{(2\pi)^2}
	d\Pi_{q}
	d\Pi_{k}
	d\Pi_{l}
	\frac{(2\pi)^4}{2E_p}
	\delta^{(4)}
	\left(
	p+q
	-k-l
	\right) \notag \\
	&|\mathcal M|^2_{\chi\chi\leftrightarrow \bm a \bm a}
	\times
	\left[
	f_{\chi_p} f_{\chi_q}
	(1+f_{a_k})(1+f_{a_l})
	-
	f_{a_k}f_{a_l}
	(1+f_{\chi_p})(1+f_{\chi_q})
	\right] .
	\label{eq:collision_full_chichi_aa}
\end{align}
The symmetry factor for the identical final state particles is understood to be included in the definition of \(\sigma_{\chi\chi\to \bm a \bm a}\).
In the Maxwell Boltzmann approximation, we neglect Bose enhancement factors and set \(1+f\simeq 1\). In this approximation, we treat only the tagged incoming dark matter particle \(\chi(p)\) as out of equilibrium, while all other particles are assumed to be described by local equilibrium distributions. The non-equilibrium distribution is written as
\begin{equation}
	f_{\chi_p}(z,p_z,p_\perp)
	=
	\mathcal A(z,p_z)f_{{\chi_p},+}^{\rm eq}(z,p).
\end{equation}
where the equilibrium distribution of the incoming dark matter in front of the wall is
\begin{equation}
	f_{{\chi_p},+}^{\rm eq}(z,p)
	=
	\exp
	\left[
	-
	\frac{
		\tilde\gamma_+
		\left(E_p-\tilde v_+p_z\right)
	}
	{T_+}
	\right].
	\label{eq:fchi_plus_eq_collision}
\end{equation}
In the absence of hydrodynamic heating, one has
\begin{equation}
	\tilde v_+=\xi_w,
	\qquad
	T_+=T_n .
\end{equation}
For the collision partner we use the approximation $f_{\chi_q}	\simeq	f_{\chi_q,+}^{\rm eq}$. The final state phase space can be expressed in terms of the annihilation cross section,
\begin{equation}
	\int
	d\Pi_{k}
	d\Pi_{l}
	(2\pi)^4
	\delta^{(4)}
	\left(
	p+q
	-k-l
	\right)
	|\mathcal M|^2_{\chi\chi\to \bm a \bm a}
	=
	4F\sigma_{\chi\chi\to \bm a \bm a},
\end{equation}
where
\begin{equation}
	F
	=
	\frac12
	\sqrt{s\left[s-4m_\chi^2(z)\right]} .
\end{equation}
\(s\equiv (p+q)^2\) is the Mandelstam invariant, corresponding to the squared center of mass energy of the incoming \(\chi\chi\) system.
Eq.~\eqref{eq:collision_full_chichi_aa} becomes
\begin{align}
	g_\chi
	\int
	\frac{dp_x dp_y}{(2\pi)^2}
	\mathbf C[f_\chi]
	=
	-g_\chi^2
	\int
	\frac{dp_x dp_y}{(2\pi)^2 2E_p}
	d\Pi_{q}
	4F \sigma_{\chi\chi\to \bm a \bm a}
	\left[
	\mathcal A(z,p_z)
	f_{\chi_p,+}^{\rm eq}
	f_{\chi_q,+}^{\rm eq}
	-
	f_{\chi_p}^{\rm eq}
	f_{\chi_q}^{\rm eq}
	\right].
	\label{eq:collision_sigma_chichi_aa}
\end{align}
To express the result in the wall frame, we write the energy of the particle in the plasma frame as
\begin{equation}
	E_p
	=
	\tilde\gamma_+
	\left(
	E_p^w-\tilde v_+ p_z^w
	\right),
	\qquad
	E_p^w
	=
	\sqrt{
		(p_x^w)^2+(p_y^w)^2+(p_z^w)^2+m_\chi^2(z)
	}.
\end{equation}

\subsection{Numerical solution of the reduced boltzmann equation and number density}

Combining the integrated Liouville operator in Eq.~\eqref{eq_liouville} with the collision term in Eq.~\eqref{eq:collision_sigma_chichi_aa}, we obtain the reduced Boltzmann equation for the non-equilibrium correction factor \(\mathcal A(z,p_z)\). This is the equation used in our numerical analysis
\begin{align}
	\left[
	\left(
	\frac{p_z}{m_\dm}
	\frac{\partial}{\partial z}
	-
	m_\dm'
	\frac{\partial}{\partial p_z}
	-
	m_\dm'
	\frac{\gp\vp}{T_+}
	\right)
	\mathcal A(z,p_z)
	\right]
	\frac{g_\dm m_\dm T_+}{2\pi\gp}
	e^{\frac{\gp}{T_+}(\vp p_z-\sqrt{p_z^2+m_\dm^2(z)})}
	=\notag \\
-g_\chi^2
\int
\frac{dp_x dp_y}{(2\pi)^2 2E_p}
d\Pi_{q}
4F \sigma_{\chi\chi\to aa}
\left[
\mathcal A(z,p_z)
f_{\chi_p,+}^{\rm eq}
f_{\chi_q,+}^{\rm eq}
-
f_{\chi_p}^{\rm eq}
f_{\chi_q}^{\rm eq}
\right]	
\label{eq:Boltzmann}
\end{align}
To solve this first order partial differential equation, we use the method of characteristics. We first define 
\begin{equation} 
	\mathcal N(z,p_z) \equiv \frac{g_\dm m_\dm T_+}{2\pi\gp} \exp\left[ \frac{\gp}{T_+} \left( \vp p_z-\sqrt{p_z^2+m_\dm^2(z)} \right) \right] . 
\end{equation} 
Then Eq.~\eqref{eq:Boltzmann} can be rewritten as 
\begin{equation} 
	\left( \frac{p_z}{m_\dm} \frac{\partial}{\partial z} - m_\dm' \frac{\partial}{\partial p_z} \right) \mathcal A 
	= m_\dm' \frac{\gp\vp}{T_+} \mathcal A - \Gamma_+(z,p_z) \mathcal A + \Gamma_{\rm eq}(z,p_z), 
	\label{eq:Boltzmann_standard_characteristic}
\end{equation}
where 
\begin{align} 
	\Gamma_+(z,p_z) &\equiv \frac{g_\chi^2}{\mathcal N(z,p_z)} \int \frac{dp_x dp_y}{(2\pi)^2 2E_p} d\Pi_q\, 4F \sigma_{\chi\chi\to \bm a \bm a} f_{\chi_p,+}^{\rm eq} f_{\chi_q,+}^{\rm eq}, \\ 
	\Gamma_{\rm eq}(z,p_z) &\equiv \frac{g_\chi^2}{\mathcal N(z,p_z)} \int \frac{dp_x dp_y}{(2\pi)^2 2E_p} d\Pi_q\, 4F \sigma_{\chi\chi\to \bm a \bm a} f_{\chi_p}^{\rm eq} f_{\chi_q}^{\rm eq}. 
\end{align} 
The characteristic curves in phase space are defined by 
\begin{equation} 
	z=z(\lambda), \qquad p_z=p_z(\lambda),
\end{equation}
with 
\begin{equation} 
	\frac{dz}{d\lambda} = \frac{p_z}{m_\dm(z)}, \qquad \frac{dp_z}{d\lambda} = - m_\dm'(z). 
	\label{eq:characteristic_curves} 
\end{equation} 
Along a characteristic curve, we define 
\begin{equation} 
	\mathcal A(\lambda) \equiv \mathcal A\bigl(z(\lambda),p_z(\lambda)\bigr). 
\end{equation}
Its total derivative is therefore 
\begin{align} 
	\frac{d\mathcal A}{d\lambda} 
	&= \frac{\partial \mathcal A}{\partial z} \frac{dz}{d\lambda} + \frac{\partial \mathcal A}{\partial p_z} \frac{dp_z}{d\lambda} \nonumber\\ 
	&= \frac{p_z}{m_\dm} \frac{\partial \mathcal A}{\partial z} - m_\dm' \frac{\partial \mathcal A}{\partial p_z}. 
\end{align} 
the Boltzmann equation reduces to the ordinary differential equation
\begin{equation} 
	\frac{d\mathcal A}{d\lambda} = \left[ m_\dm' \frac{\gp\vp}{T_+} - \Gamma_+(z,p_z) \right] \mathcal A + \Gamma_{\rm eq}(z,p_z). 
	\label{eq:A_characteristic} 
\end{equation} 
The boundary condition is fixed by assuming thermal equilibrium far in front
of the wall,
\begin{equation}
	\mathcal A(z\ll -L_w,p_z>0)=1 .
\end{equation}
We then solve the characteristic equations for particles entering the wall from the symmetric phase and obtain \(\mathcal A(z\gg L_w,p_z>0)\) deep inside the bubble. For particles moving in the opposite direction inside the bubble, we assume that they are supplied by an identical parallel wall on the other side of the bubble. This leads to
\begin{equation}
	\mathcal A(z\gg L_w,p_z)=\mathcal A(z\gg L_w,-p_z).
\end{equation}

\begin{figure}[t]
	\centering
	\begin{tabular}{@{}ccc@{}}
		\includegraphics[width=0.46\textwidth]{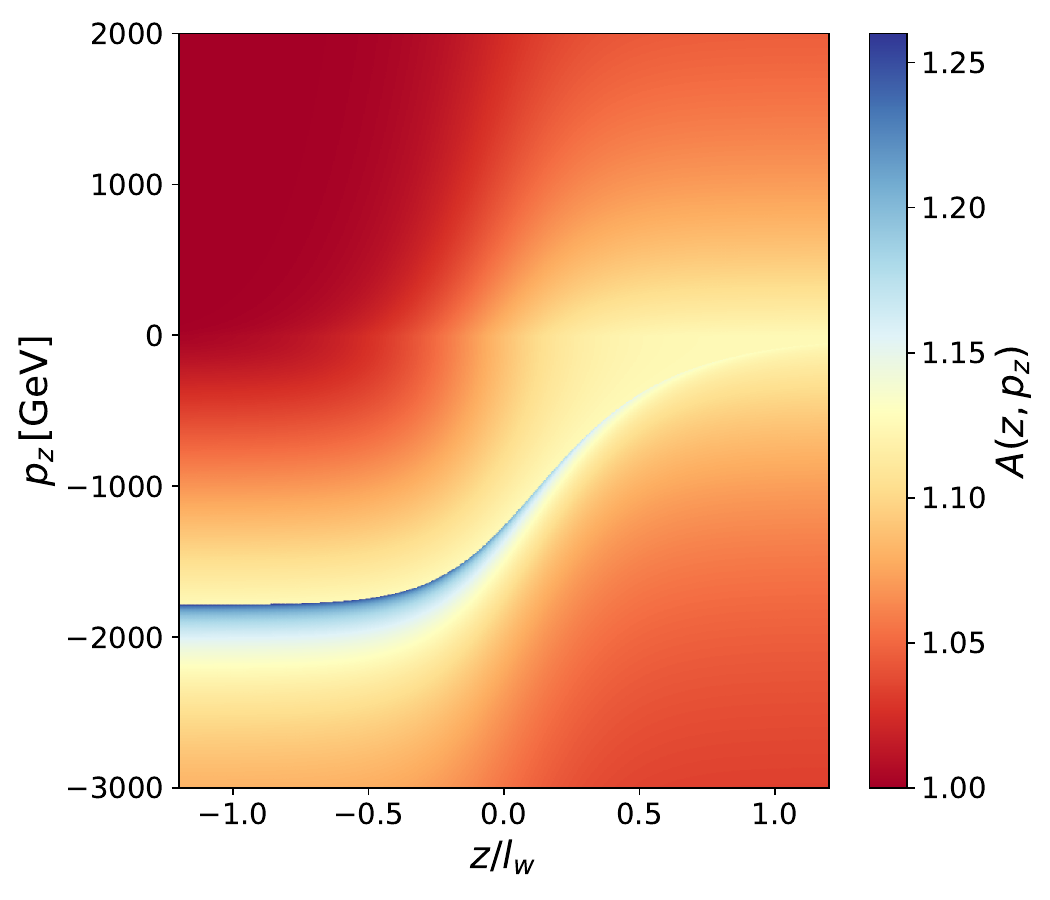} &
		\includegraphics[width=0.46\textwidth]{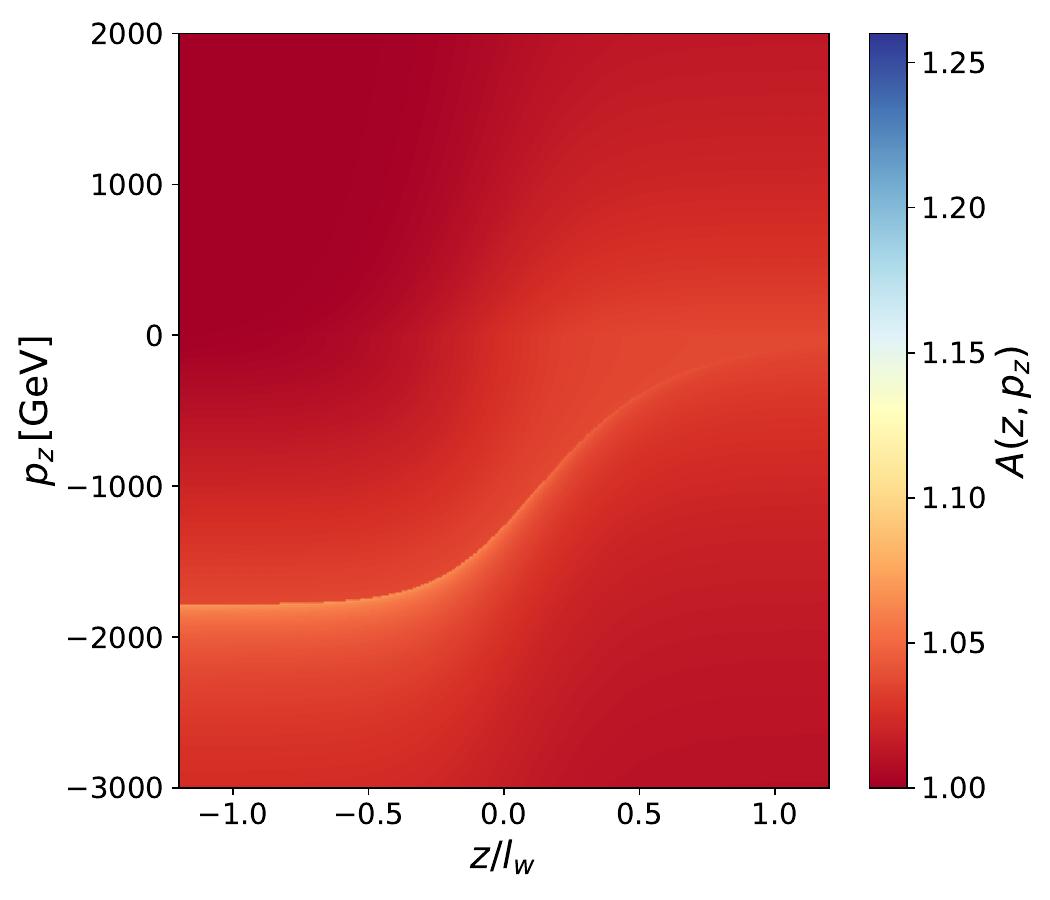} 
	\end{tabular}
	\caption{The correction factor \(\mathcal A\) for BP3 without hydrodynamic effects (left) and with hydrodynamic effects (right).}
		\label{fig:cal_A_distribution}
\end{figure}

As an illustration, Fig.~\ref{fig:cal_A_distribution} shows the numerical solution for the correction factor \(\mathcal A\) for BP3. The left panel corresponds to the result without hydrodynamic effects, while the right panel includes the hydrodynamic corrections to the temperature and velocity of the plasma in front of the wall. In most of phase space, \(\mathcal A\) remains close to unity. The main non-equilibrium correction is localized around a curved band in the \((z/l_w,p_z)\) plane. This structure originates from the spatially varying dark matter mass across the bubble wall. The mass gradient force term in the Liouville operator changes the $p_z$ of the particles, so the correction follows the corresponding characteristic trajectories in phase space. The correction is most pronounced in the region where \(p_z\) is close to the kinematic transmission threshold.
In the left panel, where hydrodynamic effects are neglected, the deviation from equilibrium is more pronounced, and a larger enhancement of \(\mathcal A\) appears near the characteristic band. After including hydrodynamic effects, as shown in the right panel, the correction becomes significantly smaller and the distribution is closer to the equilibrium limit. 
This comparison shows that hydrodynamic effects modify the reference equilibrium distribution and reduce the deviation of the non-equilibrium correction factor \(\mathcal A\) from unity.

\begin{figure}[!thbp]
	\centering
	\includegraphics[width=0.49\textwidth]{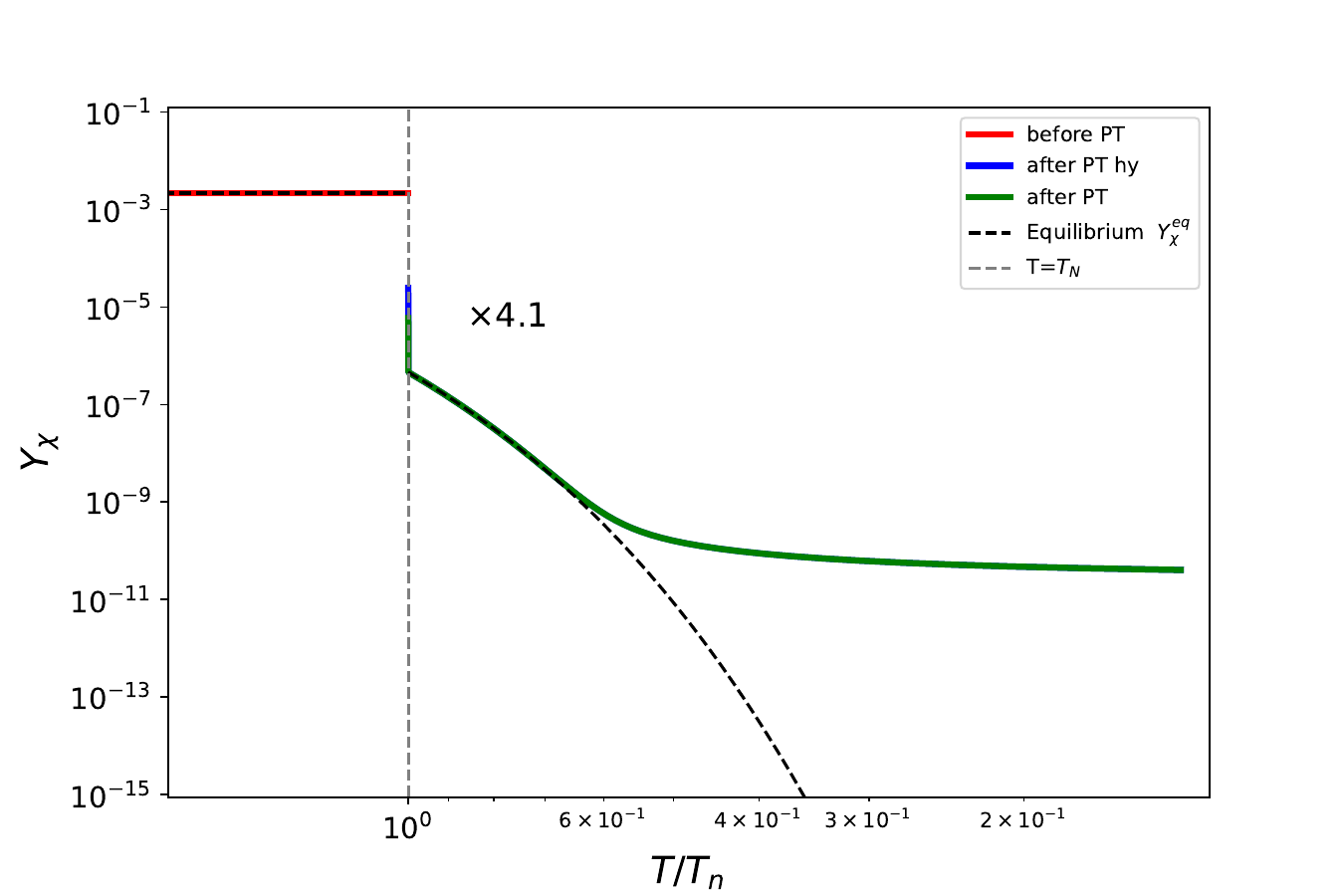} 
	\includegraphics[width=0.49\textwidth]{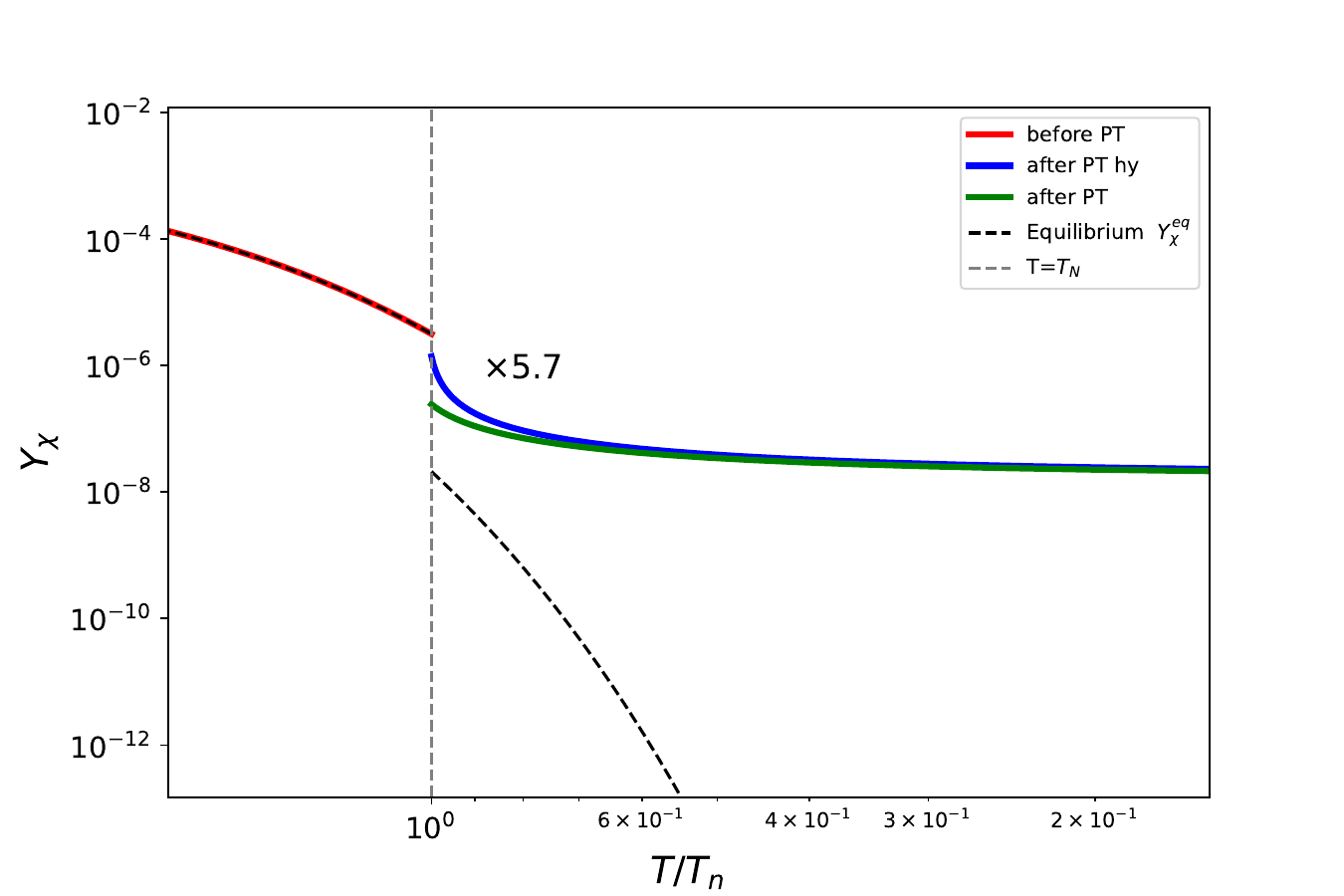} 
	\includegraphics[width=0.5\textwidth]{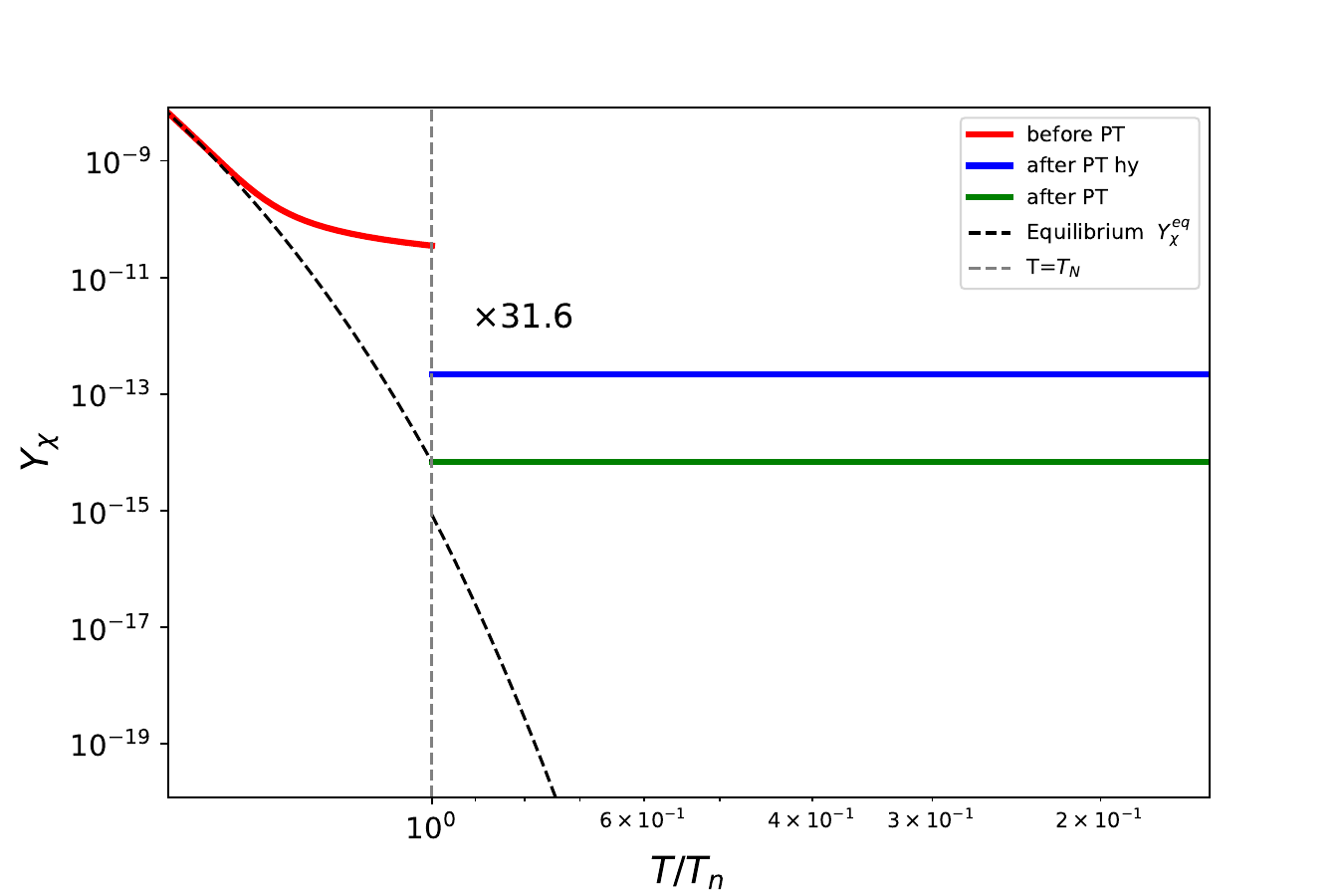}
	\caption{Numerical evolution of the dark matter abundance $Y_\dm$ as a function of $T/T_n$ for BP1 (left), BP2 (right), and BP3 (bottom). The blue and green curves show the results with and without hydrodynamic corrections, respectively. The vertical dashed line denotes the phase transition at $T=T_n$, while the black dashed curve gives the equilibrium abundance.Hydrodynamic corrections enhance the final abundance by factors of approximately $4.1$, $5.7$, and $31.6$ for BP1, BP2, and BP3, respectively.}
	\label{fig:numerical_yields}
\end{figure}

After obtaining \(\mathcal A\), we evaluate the transmitted flux using Eq.~\eqref{eq:wall_flux}, convert it to the number density through Eq.~\eqref{eq:nin_flux}, and finally obtain $Y_\dm$ using Eq.~\eqref{eq:yield_definition}. We denote the yield obtained from this numerical treatment by \(Y_{\rm num}\), in order to distinguish it from the analytic result \(Y_{\rm ana}\).
Fig.~\ref{fig:numerical_yields} shows the numerical evolution of the dark matter $Y_\dm$ for the three benchmark points as shown in Tab.~\ref{tab:analytic_numerical_comparison}.  The numerical results reproduce the thermal histories found in the analytic treatment for BP1 and BP2. 
For BP3, the large false vacuum mass strongly suppresses the dark matter abundance before nucleation, causing dark matter to freeze-out before the phase transition. Its $Y_\dm$ must be calculated numerically. For all three benchmark points, the hydrodynamic treatment increases the final filtered abundance. Quantitatively, the numerical results give $\mathcal E_{\rm hydro}\equiv Y_{\rm num}^{\rm hydro}/Y_{\rm num}^{\rm no\,hydro}\simeq 4.1$, $5.7$, and $31.6$ for BP1, BP2, and BP3, respectively. The particularly large enhancement for BP3 shows that the filtered abundance remains strongly sensitive to the plasma conditions in front of the wall even when dark matter has already frozen out before nucleation.

Tab.~\ref{tab:analytic_numerical_comparison} compares the analytic and numerical number densitys for BP1 and BP2. The analytic and numerical results are not identical, the numerical results are slightly larger than the analytic estimates. We quantify their difference by
\begin{equation}
	\delta
	\equiv
	\frac{Y_{\rm num}-Y_{\rm ana}}{Y_{\rm ana}} .
\end{equation}
Without hydrodynamic effects, the deviations are 4.7\% for BP1 and 1.2\% for BP2. After hydrodynamic effects are included, they decrease to 1.1\% and 0.2\%, respectively. These comparisons show that the analytic treatment describes the dependence of the number density on the model parameters, while the width of the wall, the mass profile, and transport in phase space account for the remaining differences. Since no analytic result is available for BP3, Tab.~\ref{tab:analytic_numerical_comparison} reports only its numerical number density.

\begin{table}[t]
	\centering
	\small
	\renewcommand{\arraystretch}{1.18}
	\setlength{\tabcolsep}{5pt}
	\begin{tabular}{ c c c c c c c c }
		\hline
		Benchmark
		& Treatment
		& \(m_\dm\,[\mathrm{GeV}]\)
		& \(m_0\,[\mathrm{GeV}]\)
		& \(Y_{\rm ana}\)
		& \(Y_{\rm num}\)
		& \(\delta\)
		& \(\mathcal E_{\rm hydro}\) \\
		\hline
		BP1
		& no hydro.
		& \(1783.81\)
		& \(0.00\)
		& \(5.65\times 10^{-6}\)
		& \(5.91\times 10^{-6}\)
		& \(4.7\%\)
		& -- \\
		BP1
		& hydro.
		& \(1783.81\)
		& \(0.00\)
		& \(2.42\times 10^{-5}\)
		& \(2.45\times 10^{-5}\)
		& \(1.1\%\)
		& \(4.1\) \\
		\hline
		BP2
		& no hydro.
		& \(2300.30\)
		& \(1452.38\)
		& \(2.40\times 10^{-7}\)
		& \(2.43\times 10^{-7}\)
		& \(1.2\%\)
		& -- \\
		BP2
		& hydro.
		& \(2300.30\)
		& \(1452.38\)
		& \(1.38\times 10^{-6}\)
		& \(1.39\times 10^{-6}\)
		& \(0.2\%\)
		& \(5.7\) \\
		\hline
		BP3
		& no hydro.
		& \(5030.30\)
		& \(4703.40\)
        & \(-\)
		& \(6.90\times 10^{-15}\)
        & \(-\)
		& --\\
	    BP3
		& hydro.
		& \(5030.30\)
		& \(4703.40\)
        & \(-\)
		& \(2.18\times 10^{-13}\)
        & \(-\)
		& \(31.6\)\\
		\hline
	\end{tabular}
	\caption{Comparison between analytic estimates and numerical results for the filtered dark matter yield, with \(\delta\equiv(Y_{\rm num}-Y_{\rm ana})/Y_{\rm ana}\). The last column gives the numerical hydrodynamic enhancement factor, $\mathcal E_{\rm hydro}\equiv Y_{\rm{num}}^{\rm hydro}/Y_{\rm{num}}^{\rm no\,hydro}$.}
	\label{tab:analytic_numerical_comparison}
\end{table}

\section{Conclusions}
\label{sec:conclusion}
In this work, we used both analytic and numerical methods to quantify hydrodynamic corrections to the entropy normalized dark matter abundance $Y_\dm$ in three scenarios with distinct abundance evolutions before the phase transition.
In the light case, dark matter remains close to equilibrium and its abundance is mainly suppressed by wall filtering. The intermediate case undergoes thermal suppression before the phase transition, while the heavy case freezes out before nucleation because of its large false vacuum mass.
For the slow wall deflagration considered here, the analytic results show that hydrodynamic effects enhance the filtered abundance, with a larger correction for heavier dark matter. This enhancement is driven predominantly by shock heating ahead of the bubble wall, which raises $T_+$ and increases the population of dark matter particles energetic enough to overcome the mass barrier. The accompanying change in the wall-frame fluid velocity is small at \(\xi_w=0.01\), but becomes increasingly important at larger wall velocities and can counteract the temperature enhancement. The numerical treatment accounts for the evolution of the dark matter distribution across a bubble wall of finite width by solving for the nonequilibrium factor $\mathcal A$.
For BP1 and BP2, the numerical and analytic values of $Y_\dm$ differ by no more than about $5\%$. After hydrodynamic effects are included, the difference decreases to approximately $1\%$ or less. For BP3, freeze out before nucleation requires both the preceding abundance evolution and the filtered abundance to be determined numerically.

\acknowledgments
We thank Zongguo Si, Wei Chao, Siyu Jiang and Shuocheng Xu for helpful comments and discussions. This work is supported by the National Natural Science Foundation of China under Grant No. 12505123, by the Project No. ZR2024MA001 supported by Shandong Provincial Natural Science Foundation.

\bibliographystyle{JHEP}
\bibliography{biblio}

@article{Griest:1989wd,
    author = "Griest, Kim and Kamionkowski, Marc",
    title = "{Unitarity Limits on the Mass and Radius of Dark Matter Particles}",
    journal = "Phys. Rev. Lett.",
    volume = "64",
    pages = "615",
    year = "1990"
}

@article{Baker:2019ndr,
    author = "Baker, Michael J. and Kopp, Joachim and Long, Andrew J.",
    title = "{Filtered Dark Matter at a First Order Phase Transition}",
    eprint = "1912.02830",
    archivePrefix = "arXiv",
    primaryClass = "hep-ph",
    doi = "10.1103/PhysRevLett.125.151102",
    journal = "Phys. Rev. Lett.",
    volume = "125",
    pages = "151102",
    year = "2020"
}

@article{Chway:2019kft,
    author = "Chway, Dongjin and Jung, Tae Hyun and Shin, Chang Sub",
    title = "{Dark matter filtering-out effect during a first-order phase transition}",
    eprint = "1912.04238",
    archivePrefix = "arXiv",
    primaryClass = "hep-ph",
    doi = "10.1103/PhysRevD.101.095019",
    journal = "Phys. Rev. D",
    volume = "101",
    pages = "095019",
    year = "2020"
}

@article{Chao:2020adk,
    author = "Chao, Wei and Li, Xiu-Fei and Wang, Lei",
    title = "{Filtered pseudo-scalar dark matter and gravitational waves from first order phase transition}",
    eprint = "2012.15113",
    archivePrefix = "arXiv",
    primaryClass = "hep-ph",
    doi = "10.1088/1475-7516/2021/06/038",
    journal = "JCAP",
    volume = "06",
    pages = "038",
    year = "2021"
}

@article{Espinosa:2010hh,
    author = "Espinosa, Jose R. and Konstandin, Thomas and No, Jose M. and Servant, Geraldine",
    title = "{Energy Budget of Cosmological First-order Phase Transitions}",
    eprint = "1004.4187",
    archivePrefix = "arXiv",
    primaryClass = "hep-ph",
    doi = "10.1088/1475-7516/2010/06/028",
    journal = "JCAP",
    volume = "06",
    pages = "028",
    year = "2010"
}

@article{Wainwright:2011kj,
    author = "Wainwright, Carroll L.",
    title = "{CosmoTransitions: Computing Cosmological Phase Transition Temperatures and Bubble Profiles with Multiple Fields}",
    eprint = "1109.4189",
    archivePrefix = "arXiv",
    primaryClass = "hep-ph",
    doi = "10.1016/j.cpc.2012.04.004",
    journal = "Comput. Phys. Commun.",
    volume = "183",
    pages = "2006",
    year = "2012"
}

@article{Goldberg:1983nd,
    author = "Goldberg, H.",
    editor = "Srednicki, M. A.",
    title = "{Constraint on the Photino Mass from Cosmology}",
    reportNumber = "NUB-2592",
    doi = "10.1103/PhysRevLett.50.1419",
    journal = "Phys. Rev. Lett.",
    volume = "50",
    pages = "1419",
    year = "1983",
    note = "[Erratum: Phys.Rev.Lett. 103, 099905 (2009)]"
}

@article{Ellis:1983ew,
    author = "Ellis, John R. and Hagelin, J. S. and Nanopoulos, Dimitri V. and Olive, Keith A. and Srednicki, M.",
    editor = "Srednicki, M. A.",
    title = "{Supersymmetric Relics from the Big Bang}",
    reportNumber = "SLAC-PUB-3171",
    doi = "10.1016/0550-3213(84)90461-9",
    journal = "Nucl. Phys. B",
    volume = "238",
    pages = "453--476",
    year = "1984"
}

@article{Jungman:1995df,
    author = "Jungman, Gerard and Kamionkowski, Marc and Griest, Kim",
    title = "{Supersymmetric dark matter}",
    eprint = "hep-ph/9506380",
    archivePrefix = "arXiv",
    reportNumber = "SU-4240-605, UCSD-PTH-95-02, IASSNS-HEP-95-14, CU-TP-677",
    doi = "10.1016/0370-1573(95)00058-5",
    journal = "Phys. Rept.",
    volume = "267",
    pages = "195--373",
    year = "1996"
}

@article{Servant:2002aq,
    author = "Servant, Geraldine and Tait, Timothy M. P.",
    title = "{Is the lightest Kaluza-Klein particle a viable dark matter candidate?}",
    eprint = "hep-ph/0206071",
    archivePrefix = "arXiv",
    reportNumber = "ANL-HEP-PR-02-032, EFI-02-74",
    doi = "10.1016/S0550-3213(02)01012-X",
    journal = "Nucl. Phys. B",
    volume = "650",
    pages = "391--419",
    year = "2003"
}

@article{Cheng:2002ej,
    author = "Cheng, Hsin-Chia and Feng, Jonathan L. and Matchev, Konstantin T.",
    title = "{Kaluza-Klein dark matter}",
    eprint = "hep-ph/0207125",
    archivePrefix = "arXiv",
    reportNumber = "EFI-02-95, UCI-TR-2002-23, UFIFT-HEP-02-21, CERN-TH-2002-157",
    doi = "10.1103/PhysRevLett.89.211301",
    journal = "Phys. Rev. Lett.",
    volume = "89",
    pages = "211301",
    year = "2002"
}

@article{Bertone:2004pz,
    author = "Bertone, Gianfranco and Hooper, Dan and Silk, Joseph",
    title = "{Particle dark matter: Evidence, candidates and constraints}",
    eprint = "hep-ph/0404175",
    archivePrefix = "arXiv",
    reportNumber = "FERMILAB-PUB-04-047-A",
    doi = "10.1016/j.physrep.2004.08.031",
    journal = "Phys. Rept.",
    volume = "405",
    pages = "279--390",
    year = "2005"
}

@article{Feng:2010gw,
    author = "Feng, Jonathan L.",
    title = "{Dark Matter Candidates from Particle Physics and Methods of Detection}",
    eprint = "1003.0904",
    archivePrefix = "arXiv",
    primaryClass = "astro-ph.CO",
    reportNumber = "UCI-TR-2009-13",
    doi = "10.1146/annurev-astro-082708-101659",
    journal = "Ann. Rev. Astron. Astrophys.",
    volume = "48",
    pages = "495--545",
    year = "2010"
}

@article{Lee:1977ua,
    author = "Lee, Benjamin W. and Weinberg, Steven",
    editor = "Srednicki, M. A.",
    title = "{Cosmological Lower Bound on Heavy Neutrino Masses}",
    reportNumber = "FERMILAB-PUB-77-041-T",
    doi = "10.1103/PhysRevLett.39.165",
    journal = "Phys. Rev. Lett.",
    volume = "39",
    pages = "165--168",
    year = "1977"
}

@article{Huang:2017kzu,
    author = "Huang, Fa Peng and Li, Chong Sheng",
    title = "{Probing the baryogenesis and dark matter relaxed in phase transition by gravitational waves and colliders}",
    eprint = "1709.09691",
    archivePrefix = "arXiv",
    primaryClass = "hep-ph",
    reportNumber = "CTPU-17-34",
    doi = "10.1103/PhysRevD.96.095028",
    journal = "Phys. Rev. D",
    volume = "96",
    number = "9",
    pages = "095028",
    year = "2017"
}

@article{Krylov:2013qe,
    author = "Krylov, E. and Levin, A. and Rubakov, V.",
    title = "{Cosmological phase transition, baryon asymmetry and dark matter Q-balls}",
    eprint = "1301.0354",
    archivePrefix = "arXiv",
    primaryClass = "hep-ph",
    doi = "10.1103/PhysRevD.87.083528",
    journal = "Phys. Rev. D",
    volume = "87",
    number = "8",
    pages = "083528",
    year = "2013"
}

@article{Jiang:2023nkj,
    author = "Jiang, Siyu and Huang, Fa Peng and Li, Chong Sheng",
    title = "{Hydrodynamic effects on the filtered dark matter produced by a first-order phase transition}",
    eprint = "2305.02218",
    archivePrefix = "arXiv",
    primaryClass = "hep-ph",
    doi = "10.1103/PhysRevD.108.063508",
    journal = "Phys. Rev. D",
    volume = "108",
    number = "6",
    pages = "063508",
    year = "2023"
}

@article{Xu:2023lkf,
    author = "Xu, Shuocheng and Zhou, Ruiyu and Cheng, Wei and Liu, Xuewen",
    title = "{Dark matter production accompanied by gravitational wave signals during cosmological phase transitions}",
    eprint = "2312.15752",
    archivePrefix = "arXiv",
    primaryClass = "hep-ph",
    doi = "10.1140/epjc/s10052-024-13046-4",
    journal = "Eur. Phys. J. C",
    volume = "84",
    number = "7",
    pages = "677",
    year = "2024"
}

@article{Baldes:2017gzw,
    author = "Baldes, Iason and Petraki, Kalliopi",
    title = "{Asymmetric thermal-relic dark matter: Sommerfeld-enhanced freeze-out, annihilation signals and unitarity bounds}",
    eprint = "1703.00478",
    archivePrefix = "arXiv",
    primaryClass = "hep-ph",
    reportNumber = "DESY-17-034, NIKHEF-2017-009",
    doi = "10.1088/1475-7516/2017/09/028",
    journal = "JCAP",
    volume = "09",
    pages = "028",
    year = "2017"
}

@article{Smirnov:2019ngs,
    author = "Smirnov, Juri and Beacom, John F.",
    title = "{TeV-Scale Thermal WIMPs: Unitarity and its Consequences}",
    eprint = "1904.11503",
    archivePrefix = "arXiv",
    primaryClass = "hep-ph",
    doi = "10.1103/PhysRevD.100.043029",
    journal = "Phys. Rev. D",
    volume = "100",
    number = "4",
    pages = "043029",
    year = "2019"
}

@article{Azatov:2021ifm,
    author = "Azatov, Aleksandr and Vanvlasselaer, Miguel and Yin, Wen",
    title = "{Dark Matter production from relativistic bubble walls}",
    eprint = "2101.05721",
    archivePrefix = "arXiv",
    primaryClass = "hep-ph",
    reportNumber = "SISSA 03/2021/FISI",
    doi = "10.1007/JHEP03(2021)288",
    journal = "JHEP",
    volume = "03",
    pages = "288",
    year = "2021"
}

@article{Azatov:2022tii,
    author = "Azatov, Aleksandr and Barni, Giulio and Chakraborty, Sabyasachi and Vanvlasselaer, Miguel and Yin, Wen",
    title = "{Ultra-relativistic bubbles from the simplest Higgs portal and their cosmological consequences}",
    eprint = "2207.02230",
    archivePrefix = "arXiv",
    primaryClass = "hep-ph",
    reportNumber = "SISSA 12/2022/FISI TU-1157",
    doi = "10.1007/JHEP10(2022)017",
    journal = "JHEP",
    volume = "10",
    pages = "017",
    year = "2022"
}

@article{Baldes:2022oev,
    author = "Baldes, Iason and Gouttenoire, Yann and Sala, Filippo",
    title = "{Hot and heavy dark matter from a weak scale phase transition}",
    eprint = "2207.05096",
    archivePrefix = "arXiv",
    primaryClass = "hep-ph",
    reportNumber = "ULB-TH/22-12",
    doi = "10.21468/SciPostPhys.14.3.033",
    journal = "SciPost Phys.",
    volume = "14",
    number = "3",
    pages = "033",
    year = "2023"
}

@article{Xiao:2022oaq,
    author = "Xiao, Yang and Yang, Jin Min and Zhang, Yang",
    title = "{Dilution of dark matter relic density in singlet extension models}",
    eprint = "2207.14519",
    archivePrefix = "arXiv",
    primaryClass = "hep-ph",
    doi = "10.1007/JHEP02(2023)008",
    journal = "JHEP",
    volume = "02",
    pages = "008",
    year = "2023"
}

@article{Ai:2021kak,
    author = "Ai, Wen-Yuan and Garbrecht, Bjorn and Tamarit, Carlos",
    title = "{Bubble wall velocities in local equilibrium}",
    eprint = "2109.13710",
    archivePrefix = "arXiv",
    primaryClass = "hep-ph",
    reportNumber = "CP3-21-53, TUM-HEP-1365-21",
    doi = "10.1088/1475-7516/2022/03/015",
    journal = "JCAP",
    volume = "03",
    number = "03",
    pages = "015",
    year = "2022"
}

@article{Ai:2023see,
    author = "Ai, Wen-Yuan and Laurent, Benoit and van de Vis, Jorinde",
    title = "{Model-independent bubble wall velocities in local thermal equilibrium}",
    eprint = "2303.10171",
    archivePrefix = "arXiv",
    primaryClass = "astro-ph.CO",
    reportNumber = "KCL-PH-TH/2023-19",
    doi = "10.1088/1475-7516/2023/07/002",
    journal = "JCAP",
    volume = "07",
    pages = "002",
    year = "2023"
}

@article{Cline:2020jre,
    author = "Cline, James M. and Kainulainen, Kimmo",
    title = "{Electroweak baryogenesis at high bubble wall velocities}",
    eprint = "2001.00568",
    archivePrefix = "arXiv",
    primaryClass = "hep-ph",
    reportNumber = "CERN-TH-2019-227",
    doi = "10.1103/PhysRevD.101.063525",
    journal = "Phys. Rev. D",
    volume = "101",
    number = "6",
    pages = "063525",
    year = "2020"
}

@article{Laurent:2020gpg,
    author = "Laurent, Benoit and Cline, James M.",
    title = "{Fluid equations for fast-moving electroweak bubble walls}",
    eprint = "2007.10935",
    archivePrefix = "arXiv",
    primaryClass = "hep-ph",
    doi = "10.1103/PhysRevD.102.063516",
    journal = "Phys. Rev. D",
    volume = "102",
    number = "6",
    pages = "063516",
    year = "2020"
}

@article{Cline:2021iff,
    author = "Cline, James M. and Friedlander, Avi and He, Dong-Ming and Kainulainen, Kimmo and Laurent, Benoit and Tucker-Smith, David",
    title = "{Baryogenesis and gravity waves from a UV-completed electroweak phase transition}",
    eprint = "2102.12490",
    archivePrefix = "arXiv",
    primaryClass = "hep-ph",
    doi = "10.1103/PhysRevD.103.123529",
    journal = "Phys. Rev. D",
    volume = "103",
    number = "12",
    pages = "123529",
    year = "2021"
}

@article{Lewicki:2021pgr,
    author = "Lewicki, Marek and Merchand, Marco and Zych, Mateusz",
    title = "{Electroweak bubble wall expansion: gravitational waves and baryogenesis in Standard Model-like thermal plasma}",
    eprint = "2111.02393",
    archivePrefix = "arXiv",
    primaryClass = "astro-ph.CO",
    doi = "10.1007/JHEP02(2022)017",
    journal = "JHEP",
    volume = "02",
    pages = "017",
    year = "2022"
}

@article{Giese:2020rtr,
    author = "Giese, Felix and Konstandin, Thomas and van de Vis, Jorinde",
    title = "{Model-independent energy budget of cosmological first-order phase transitions{\textemdash}A sound argument to go beyond the bag model}",
    eprint = "2004.06995",
    archivePrefix = "arXiv",
    primaryClass = "astro-ph.CO",
    reportNumber = "DESY-20-064",
    doi = "10.1088/1475-7516/2020/07/057",
    journal = "JCAP",
    volume = "07",
    number = "07",
    pages = "057",
    year = "2020"
}

@article{Giese:2020znk,
    author = "Giese, Felix and Konstandin, Thomas and Schmitz, Kai and van de Vis, Jorinde",
    title = "{Model-independent energy budget for LISA}",
    eprint = "2010.09744",
    archivePrefix = "arXiv",
    primaryClass = "astro-ph.CO",
    reportNumber = "DESY-20-173, DESY 20-173, CERN-TH-2020-170",
    doi = "10.1088/1475-7516/2021/01/072",
    journal = "JCAP",
    volume = "01",
    pages = "072",
    year = "2021"
}

@article{Leitao:2014pda,
    author = "Leitao, Leonardo and Megevand, Ariel",
    title = "{Hydrodynamics of phase transition fronts and the speed of sound in the plasma}",
    eprint = "1410.3875",
    archivePrefix = "arXiv",
    primaryClass = "hep-ph",
    doi = "10.1016/j.nuclphysb.2014.12.008",
    journal = "Nucl. Phys. B",
    volume = "891",
    pages = "159--199",
    year = "2015"
}

@article{Wang:2023jto,
    author = "Wang, Xiao and Tian, Chi and Huang, Fa Peng",
    title = "{Model-dependent analysis method for energy budget of the cosmological first-order phase transition}",
    eprint = "2301.12328",
    archivePrefix = "arXiv",
    primaryClass = "hep-ph",
    doi = "10.1088/1475-7516/2023/07/006",
    journal = "JCAP",
    volume = "07",
    pages = "006",
    year = "2023"
}

@article{Wang:2022lyd,
    author = "Wang, Shao-Jiang and Yuwen, Zi-Yan",
    title = "{The energy budget of cosmological first-order phase transitions beyond the bag equation of state}",
    eprint = "2206.01148",
    archivePrefix = "arXiv",
    primaryClass = "hep-ph",
    doi = "10.1088/1475-7516/2022/10/047",
    journal = "JCAP",
    volume = "10",
    pages = "047",
    year = "2022"
}

@article{Si:2025vdt,
    author = "Si, Zongguo and Wang, Hongxin and Wang, Lei and Xiao, Yang and Zhang, Yang",
    title = "{The bubble wall velocity in local thermal equilibrium and energy budget with full effective potential}",
    eprint = "2505.19584",
    archivePrefix = "arXiv",
    primaryClass = "hep-ph",
    doi = "10.1007/JHEP09(2025)029",
    journal = "JHEP",
    volume = "09",
    pages = "029",
    year = "2025"
}

@article{Tenkanen:2022tly,
    author = "Tenkanen, Tuomas V. I. and van de Vis, Jorinde",
    title = "{Speed of sound in cosmological phase transitions and effect on gravitational waves}",
    eprint = "2206.01130",
    archivePrefix = "arXiv",
    primaryClass = "hep-ph",
    reportNumber = "NORDITA 2022-031, DESY-22-091",
    doi = "10.1007/JHEP08(2022)302",
    journal = "JHEP",
    volume = "08",
    pages = "302",
    year = "2022"
}

@article{Tian:2024ysd,
    author = "Tian, Chi and Wang, Xiao and Bal{\'a}zs, Csaba",
    title = "{Gravitational waves from cosmological first-order phase transitions with precise hydrodynamics}",
    eprint = "2409.14505",
    archivePrefix = "arXiv",
    primaryClass = "hep-ph",
    doi = "10.1140/epjc/s10052-025-14826-2",
    journal = "Eur. Phys. J. C",
    volume = "85",
    number = "10",
    pages = "1091",
    year = "2025"
}

@article{Wang:2021jtw,
  author        = {Wang, Xiao and Huang, Fa Peng and Li, Yongping},
  title         = {{Sound velocity effects on the phase transition gravitational wave spectrum in the sound shell model}},
  journal       = {Phys. Rev. D},
  volume        = {105},
  number        = {10},
  pages         = {103513},
  year          = {2022},
  eprint        = {2112.14650},
  archivePrefix = {arXiv},
  primaryClass  = {astro-ph.CO},
  doi           = {10.1103/PhysRevD.105.103513}
}

@article{Barger:2008jx,
    author = "Barger, Vernon and Langacker, Paul and McCaskey, Mathew and Ramsey-Musolf, Michael and Shaughnessy, Gabe",
    title = "{Complex Singlet Extension of the Standard Model}",
    eprint = "0811.0393",
    archivePrefix = "arXiv",
    primaryClass = "hep-ph",
    reportNumber = "MADPH-08-1516, NUHEP-TH-08-06, ANL-HEP-PR-08-58, NPAC-08-21",
    doi = "10.1103/PhysRevD.79.015018",
    journal = "Phys. Rev. D",
    volume = "79",
    pages = "015018",
    year = "2009"
}

@article{Gross:2017dan,
    author = "Gross, Christian and Lebedev, Oleg and Toma, Takashi",
    title = "{Cancellation Mechanism for Dark-Matter{\textendash}Nucleon Interaction}",
    eprint = "1708.02253",
    archivePrefix = "arXiv",
    primaryClass = "hep-ph",
    reportNumber = "HIP-2017-20-TH, TUM-HEP-1091-17, HIP-2017-20/TH, TUM-HEP/1091/17",
    doi = "10.1103/PhysRevLett.119.191801",
    journal = "Phys. Rev. Lett.",
    volume = "119",
    number = "19",
    pages = "191801",
    year = "2017"
}

@article{Kannike:2019wsn,
    author = "Kannike, Kristjan and Raidal, Martti",
    title = "{Phase Transitions and Gravitational Wave Tests of Pseudo-Goldstone Dark Matter in the Softly Broken U(1) Scalar Singlet Model}",
    eprint = "1901.03333",
    archivePrefix = "arXiv",
    primaryClass = "hep-ph",
    doi = "10.1103/PhysRevD.99.115010",
    journal = "Phys. Rev. D",
    volume = "99",
    number = "11",
    pages = "115010",
    year = "2019"
}

@article{Kannike:2019mzk,
    author = "Kannike, Kristjan and Loos, Kaius and Raidal, Martti",
    title = "{Gravitational wave signals of pseudo-Goldstone dark matter in the $\mathbb{Z}_{3}$ complex singlet model}",
    eprint = "1907.13136",
    archivePrefix = "arXiv",
    primaryClass = "hep-ph",
    doi = "10.1103/PhysRevD.101.035001",
    journal = "Phys. Rev. D",
    volume = "101",
    number = "3",
    pages = "035001",
    year = "2020"
}

@article{Alanne:2020jwx,
    author = "Alanne, Tommi and Benincasa, Nico and Heikinheimo, Matti and Kannike, Kristjan and Keus, Venus and Koivunen, Niko and Tuominen, Kimmo",
    title = "{Pseudo-Goldstone dark matter: gravitational waves and direct-detection blind spots}",
    eprint = "2008.09605",
    archivePrefix = "arXiv",
    primaryClass = "hep-ph",
    reportNumber = "HIP-2020-24/TH",
    doi = "10.1007/JHEP10(2020)080",
    journal = "JHEP",
    volume = "10",
    pages = "080",
    year = "2020"
}

@article{Coito:2021fgo,
    author = "Coito, Leonardo and Faubel, Carlos and Herrero-Garcia, Juan and Santamaria, Arcadi",
    title = "{Dark matter from a complex scalar singlet: the role of dark CP and other discrete symmetries}",
    eprint = "2106.05289",
    archivePrefix = "arXiv",
    primaryClass = "hep-ph",
    reportNumber = "FTUV-21-0608.9540, IFIC/21-20",
    doi = "10.1007/JHEP11(2021)202",
    journal = "JHEP",
    volume = "11",
    pages = "202",
    year = "2021"
}

@article{Dolan:1973qd,
    author = "Dolan, L. and Jackiw, R.",
    title = "{Symmetry Behavior at Finite Temperature}",
    reportNumber = "MIT-CTP-406",
    doi = "10.1103/PhysRevD.9.3320",
    journal = "Phys. Rev. D",
    volume = "9",
    pages = "3320--3341",
    year = "1974"
}

@inproceedings{Quiros:1999jp,
    author = "Quiros, Mariano",
    title = "{Finite temperature field theory and phase transitions}",
    booktitle = "{ICTP Summer School in High-Energy Physics and Cosmology}",
    eprint = "hep-ph/9901312",
    archivePrefix = "arXiv",
    reportNumber = "IEM-FT-187-99",
    pages = "187--259",
    month = "1",
    year = "1999"
}

@article{Coleman:1977py,
    author = "Coleman, Sidney R.",
    title = "{The Fate of the False Vacuum. 1. Semiclassical Theory}",
    reportNumber = "HUTP-77-A004",
    doi = "10.1103/PhysRevD.15.2929",
    journal = "Phys. Rev. D",
    volume = "15",
    pages = "2929--2936",
    year = "1977",
    note = "[Erratum: Phys.Rev.D 16, 1248 (1977)]"
}

@article{Callan:1977pt,
    author = "Callan, Jr., Curtis G. and Coleman, Sidney R.",
    title = "{The Fate of the False Vacuum. 2. First Quantum Corrections}",
    reportNumber = "HUTP-77-A032",
    doi = "10.1103/PhysRevD.16.1762",
    journal = "Phys. Rev. D",
    volume = "16",
    pages = "1762--1768",
    year = "1977"
}

@article{Linde:1981zj,
    author = "Linde, Andrei D.",
    title = "{Decay of the False Vacuum at Finite Temperature}",
    reportNumber = "LEBEDEV-81-265",
    doi = "10.1016/0550-3213(83)90072-X",
    journal = "Nucl. Phys. B",
    volume = "216",
    pages = "421",
    year = "1983",
    note = "[Erratum: Nucl.Phys.B 223, 544 (1983)]"
}

@article{Steinhardt:1981ct,
    author = "Steinhardt, Paul Joseph",
    title = "{Relativistic Detonation Waves and Bubble Growth in False Vacuum Decay}",
    reportNumber = "UPR-0181T",
    doi = "10.1103/PhysRevD.25.2074",
    journal = "Phys. Rev. D",
    volume = "25",
    pages = "2074",
    year = "1982"
}

@article{Kurki-Suonio:1995rrv,
    author = "Kurki-Suonio, H. and Laine, M.",
    title = "{Supersonic deflagrations in cosmological phase transitions}",
    eprint = "hep-ph/9501216",
    archivePrefix = "arXiv",
    reportNumber = "HU-TFT-95-3",
    doi = "10.1103/PhysRevD.51.5431",
    journal = "Phys. Rev. D",
    volume = "51",
    pages = "5431--5437",
    year = "1995"
}

@article{Kurki-Suonio:1995yaf,
    author = "Kurki-Suonio, H. and Laine, M.",
    title = "{On bubble growth and droplet decay in cosmological phase transitions}",
    eprint = "hep-ph/9512202",
    archivePrefix = "arXiv",
    reportNumber = "HU-TFT-95-71, HD-THEP-95-52",
    doi = "10.1103/PhysRevD.54.7163",
    journal = "Phys. Rev. D",
    volume = "54",
    pages = "7163--7171",
    year = "1996"
}

@article{Profumo:2007wc,
    author = "Profumo, Stefano and Ramsey-Musolf, Michael J. and Shaughnessy, Gabe",
    title = "{Singlet Higgs phenomenology and the electroweak phase transition}",
    eprint = "0705.2425",
    archivePrefix = "arXiv",
    primaryClass = "hep-ph",
    reportNumber = "CALTECH-MAP-333, MADPH-07-1489",
    doi = "10.1088/1126-6708/2007/08/010",
    journal = "JHEP",
    volume = "08",
    pages = "010",
    year = "2007"
}

@article{Cline:2012hg,
    author = "Cline, James M. and Kainulainen, Kimmo",
    title = "{Electroweak baryogenesis and dark matter from a singlet Higgs}",
    eprint = "1210.4196",
    archivePrefix = "arXiv",
    primaryClass = "hep-ph",
    doi = "10.1088/1475-7516/2013/01/012",
    journal = "JCAP",
    volume = "01",
    pages = "012",
    year = "2013"
}

@article{Gondolo:1990dk,
    author = "Gondolo, Paolo and Gelmini, Graciela",
    title = "{Cosmic abundances of stable particles: Improved analysis}",
    reportNumber = "UCLA-90-TEP-68",
    doi = "10.1016/0550-3213(91)90438-4",
    journal = "Nucl. Phys. B",
    volume = "360",
    pages = "145--179",
    year = "1991"
}

@article{Alloul:2013bka,
    author = "Alloul, Adam and Christensen, Neil D. and Degrande, C{\'e}line and Duhr, Claude and Fuks, Benjamin",
    title = "{FeynRules  2.0 - A complete toolbox for tree-level phenomenology}",
    eprint = "1310.1921",
    archivePrefix = "arXiv",
    primaryClass = "hep-ph",
    reportNumber = "CERN-PH-TH-2013-239, MCNET-13-14, IPPP-13-71, DCPT-13-142, PITT-PACC-1308",
    doi = "10.1016/j.cpc.2014.04.012",
    journal = "Comput. Phys. Commun.",
    volume = "185",
    pages = "2250--2300",
    year = "2014"
}

@article{Shtabovenko:2016sxi,
    author = "Shtabovenko, Vladyslav and Mertig, Rolf and Orellana, Frederik",
    title = "{New Developments in FeynCalc 9.0}",
    eprint = "1601.01167",
    archivePrefix = "arXiv",
    primaryClass = "hep-ph",
    reportNumber = "TUM-EFT-71-15",
    doi = "10.1016/j.cpc.2016.06.008",
    journal = "Comput. Phys. Commun.",
    volume = "207",
    pages = "432--444",
    year = "2016"
}

@article{Shtabovenko:2020gxv,
    author = "Shtabovenko, Vladyslav and Mertig, Rolf and Orellana, Frederik",
    title = "{FeynCalc 9.3: New features and improvements}",
    eprint = "2001.04407",
    archivePrefix = "arXiv",
    primaryClass = "hep-ph",
    reportNumber = "P3H-20-002, TTP19-020, TUM-EFT 130/19",
    doi = "10.1016/j.cpc.2020.107478",
    journal = "Comput. Phys. Commun.",
    volume = "256",
    pages = "107478",
    year = "2020"
}

@article{Joyce:1994zn,
    author = "Joyce, Michael and Prokopec, Tomislav and Turok, Neil",
    title = "{Nonlocal electroweak baryogenesis. Part 1: Thin wall regime}",
    eprint = "hep-ph/9410281",
    archivePrefix = "arXiv",
    reportNumber = "PUPT-1495, PUP-TH-1495 (1994)",
    doi = "10.1103/PhysRevD.53.2930",
    journal = "Phys. Rev. D",
    volume = "53",
    pages = "2930--2957",
    year = "1996"
}

@article{Moore:1995si,
    author = "Moore, Guy D. and Prokopec, Tomislav",
    title = "{How fast can the wall move? A Study of the electroweak phase transition dynamics}",
    eprint = "hep-ph/9506475",
    archivePrefix = "arXiv",
    reportNumber = "PUPT-1544, PUP-TH-1544, LANCS-TH-9517",
    doi = "10.1103/PhysRevD.52.7182",
    journal = "Phys. Rev. D",
    volume = "52",
    pages = "7182--7204",
    year = "1995"
}

@article{Cline:2000nw,
    author = "Cline, James M. and Joyce, Michael and Kainulainen, Kimmo",
    title = "{Supersymmetric electroweak baryogenesis}",
    eprint = "hep-ph/0006119",
    archivePrefix = "arXiv",
    reportNumber = "MCGILL-00-15, NORDITA-2000-38-HE, LPT-ORSAY-00-46",
    doi = "10.1088/1126-6708/2000/07/018",
    journal = "JHEP",
    volume = "07",
    pages = "018",
    year = "2000"
}

@article{Wada:2026iux,
    author = "Wada, Juntaro",
    title = "{Hydrodynamics of Filtered Dark Matter: A Two-Component Approach}",
    eprint = "2604.17383",
    archivePrefix = "arXiv",
    primaryClass = "hep-ph",
    reportNumber = "TU-1303",
    month = "4",
    year = "2026"
}

\end{document}